\documentclass[11pt,noindent]{article}
\usepackage{graphics,cite,amssymb,epsfig,float}
\usepackage[usenames,dvips]{color}
\usepackage{amsmath, mathbbol}
\usepackage{amsfonts,xfrac,slashed}
\usepackage{pstricks}
\usepackage{a4,graphicx}
\usepackage{pst-plot}
\usepackage{fancybox}
\usepackage{amsfonts}

\usepackage{ulem}

\usepackage{array}
\usepackage{multirow}
\def\lsim{\;\raise0.3ex\hbox{$<$\kern-0.75em\raise-1.1ex\hbox{$\sim$}}\;}
\def\gsim{\;\raise0.3ex\hbox{$>$\kern-0.75em\raise-1.1ex\hbox{$\sim$}}\;}

\def\ben{\begin{enumerate}}  \def\een{\end{enumerate}}
\def\bit{\begin{itemize}}    \def\eit{\end{itemize}}
\def\beq{\begin{equation}}   \def\eeq{\end{equation}}
\def\ba{\begin{array}}       \def\ea{\end{array}}
\def\bea{\begin{eqnarray}}   \def\eea{\end{eqnarray}}

\newcolumntype{C}{ >{\centering\arraybackslash} m{4cm} }

\begin{document}

\setcounter{footnote}{0}
\vspace*{-1.5cm}
\begin{flushright}
LPT Orsay 16-86 \\
PCCF RI 16-08\\
IFT-UAM/CSIC-16-138\\
FTUAM-16-47\\

\vspace*{2mm}
\today								
\end{flushright}
\begin{center}
\vspace*{5mm}

\vspace{1cm}
{\Large\bf 
In-flight cLFV conversion: $\pmb{e-\mu}$, $\pmb{e-\tau}$ and 
$\pmb{\mu-\tau}$
in minimal extensions of the Standard Model with sterile fermions}
\vspace*{0.8cm}

{\bf A. Abada$^{a}$, V. De Romeri$^{b}$, J. Orloff$^{c}$ and
  A.M. Teixeira$^{c}$ }  
   
\vspace*{.5cm} 
$^{a}$Laboratoire de Physique Th\'eorique, CNRS, \\
Univ. Paris-Sud, Universit\'e Paris-Saclay, 91405 Orsay, France
§\vspace*{.2cm} 

$^{b}$ Departamento de F\'{\i}sica Te\'orica and Instituto de 
F\'{\i}sica Te\'orica, IFT-UAM/CSIC,\\
Universidad Aut\'onoma de Madrid, Cantoblanco, 28049 Madrid, Spain
\vspace*{.2cm} 

$^{c}$ Laboratoire de Physique Corpusculaire, CNRS/IN2P3 -- UMR 6533,\\ 
Campus des C\'ezeaux, 4 Av. Blaise Pascal, F-63178 Aubi\`ere Cedex, France
\end{center}

\vspace*{6mm}
\begin{abstract}
We revisit charged lepton flavour in-flight conversions, in which a
beam of electrons or muons is directed onto a fixed target, 
$e + N \to \mu +N$, $e + N \to \tau +N$ and $\mu + N \to \tau +N$, 
focusing on elastic interactions with a nucleus $N$.
After a general discussion of this observable, 
we carry a full phenomenological analysis 
in the framework of minimal Standard Model extensions via sterile
neutrinos, with a strong emphasis on the r\^ole of the increasingly more
stringent constraints arising from other (low-energy) charged lepton
flavour violation observables.
Despite the potential interest of this observable, in particular in
the light of certain upcoming facilities with the 
capability of very intense lepton beams,  
our study suggests that due to current bounds on
three-body decays ($\ell_i \to 3 \ell_j$) and $\mu-e$ conversion in
Nuclei, the expected number of conversions in such a minimal framework
is dramatically reduced. An experimental observation of such a
conversion would thus signal the presence of another source of flavour
violation, possibly at tree-level.

\end{abstract}
\vspace*{6mm}

\section{Introduction}

The quest for a Standard Model (SM) extension 
capable of addressing its several observational  
caveats has fuelled intensive experimental 
searches, encompassing high-energy
colliders, high-intensity facilities, as well as numerous 
astroparticle and cosmological searches. 

So far, no direct evidence for the new states has been unveiled in
collider searches, and this has in turn intensified the interest
for the so-called indirect searches, in which very rare
processes, strongly suppressed or even forbidden in the SM, are looked for. 
Among the many observables that are being studied and explored, those
signaling the violation of lepton flavour are powerful probes of New
Physics (NP), sensitive to new scales often lying well beyond collider
reach. Numerous processes are currently being searched for in
high-intensity facilities, and these include charged lepton flavour violating 
(cLFV) radiative decays, three-body decays and nuclear-assisted
transitions; likewise, a vast array of rare transitions and decays
is being looked for at high-energy colliders\footnote{Other rare
  processes, such as those violating lepton flavour universality (LFUV), or
  total lepton number (LNV), can also emerge in relation with cLFV
  transitions and decays.}.  
The current bounds are already impressive, and many running and/or
upcoming experiments (as is the case of MEG II, Mu3e, Mu2e, COMET and
LHCb) should improve them in the near future. In 
Table~\ref{table:cLFV:bounds} we summarise the present 
experimental bounds and future sensitivities for several 
radiative and 3-body cLFV decays (which will be relevant for our 
subsequent discussion).  

\begin{table}[h!]
\centering
\begin{tabular}{|c|c|c|}
\hline
cLFV Process & Present Bound & Future Sensitivity  \\
\hline
    $\mu \rightarrow  e \gamma$ & $4.2\times
10^{-13}$~\cite{TheMEG:2016wtm}  & $6\times
10^{-14}$~\cite{Baldini:2013ke} \\ 
    $\tau \to e \gamma$ & $3.3 \times 10^{-8}$~\cite{Aubert:2009ag}& $
\sim3\times10^{-9}$~\cite{Aushev:2010bq}\\ 
    $\tau \to \mu \gamma$ & $4.4 \times 10^{-8}$~\cite{Aubert:2009ag}&
$ \sim3\times10^{-9}$~\cite{Aushev:2010bq} \\ 
\hline
    $\mu \rightarrow e e e$ &  $1.0 \times
10^{-12}$~\cite{Bellgardt:1987du} &
$\sim10^{-16}$~\cite{Blondel:2013ia} \\ 
    $\tau \rightarrow \mu \mu \mu$ &
$2.1\times10^{-8}$~\cite{Hayasaka:2010np} & $\sim
10^{-9}$~\cite{Aushev:2010bq} \\ 
    $\tau \rightarrow e e e$ &
$2.7\times10^{-8}$~\cite{Hayasaka:2010np} &  $\sim
10^{-9}$~\cite{Aushev:2010bq} \\ 
\hline
$\mu - e$ &
$7\times10^{-13}$ (Au)~\cite{Bertl:2006up} &  
\begin{tabular}{c}
$\sim
10^{-14}$ (SiC)~\cite{Aoki:2012zza}  \\
$\sim
10^{-17}$ (Al)~\cite{Carey:2008zz,Cui:2009zz,Kuno:2013mha}  
\end{tabular}
\\ 
\hline
\end{tabular}
\caption{Current experimental bounds and future sensitivities for
 several cLFV processes, which are considered 
in this study.}
\label{table:cLFV:bounds}
\end{table}

The probing power of cLFV has been at the source of an increasing
interest for these processes, leading to further explorations
of already existing observables, or to the study of  
new ones. This was the case of the Coulomb enhanced decays of a muonic
atom into two electrons~\cite{Koike:2010xr,Uesaka:2016vfy}, 
or the lepton flavour and lepton number
violation $\mu^- - e^+$ conversion in
Nuclei~\cite{Littenberg:2000fg,Geib:2016atx,Berryman:2016slh,Geib:2016daa}. 

In the wake of the discovery of $\nu_\mu - \nu_\tau$ oscillations -
and of large mixing in the neutral lepton sector - 
the study of cLFV $\tau$ lepton production in $\mu + N \to \tau + N$
(with $N$ denoting a generic nucleon) at high 
energies~\cite{Gninenko:2001id} was originally proposed.
The experimental signature  for
the $\mu-\tau$ cLFV in-flight conversion would be that of a final
state composed by a single muon (the tau, despite its large energy,
rapidly decaying, $\tau \to \mu \nu \nu$), with a dramatic loss in
energy when compared to that of the primary muon beam - the
energy loss on target corresponding to the 
production and subsequent decay of the heavier lepton. 
First studies also focused on the
quasi-elastic in-flight conversion, 
due to the simpler final state topology and to the associated background.
The possibility of having 
high-intensity (and sufficiently energetic) 
muon beams (for instance at muon and future neutrino factories)
further fuelled the interest for such cLFV observables; as argued 
in~\cite{Sher:2003vi}, a 50~GeV muon beam, with an expected intensity
of $10^{20}$ muons on target per year could lead to a significant number
of $\mu + N \to \tau + N$ events. The original estimation was based on
an effective approach, and preceeded the recent stringent bounds on
cLFV transitions (many of them collected in
Table~\ref{table:cLFV:bounds}). 

Other pioneering studies of in-flight cLFV conversion focused on leptoquark
models~\cite{Gninenko:2001id,Gonderinger:2010yn}, also highlighting
the potential of flavour violating constructions such as $R$-parity 
($R_P$)-violating supersymmetry, or flavour-violating Higgs interactions.  
Following the model-independent approach of~\cite{Sher:2003vi}, 
the prospects of supersymmetric extensions of the SM for 
$\mu + N \to \tau + X_\text{final}$ were
discussed~\cite{Kanemura:2004jt} in the deep-inelastic scattering (DIS)
regime, as were those of  
low-energy electron-nucleus scattering to probe $e \to \mu$ 
conversion~\cite{Blazek:2004cg}. 
In {Ref.}~\cite{Diener:2004kq}, the impact of massive neutrinos was first
considered; contributions arising in the framework of a typical type I
seesaw (albeit for low right-handed neutrino masses) were found to
enhance those emerging from the presence of three light massive
Dirac neutrinos, assuming a
CKM-like lepton mixing, by as much as twenty orders of magnitude 
in the case of dominating photon contributions.
Other studies in the DIS regime focused on cLFV conversions induced by
``unparticles''~\cite{Bolanos:2012zd}. 
Recent analyses, again based on an effective-Lagrangian approach, 
included a detailed discussion of the process' kinematics and hadronic
contributions~\cite{Liao:2015hqu}. Associated experimental issues
(including a brief overview of backgrounds), and 
future prospects were discussed in~\cite{Bernstein:2013hba}. 

In view of recent phenomenological
and experimental developments, which have led to increasingly severe
bounds on the scale of NP mediators and to strong constraints on the
strength of possible cLFV couplings, a re-analysis of the
in-flight cLFV conversion - and its potential impact on SM extensions
- is clearly justified. Expected experimental
prospects (such as the capability of high-intensity, high-energy
muon beams~\cite{Kanemura:2004jt,SPS:NA64}, or a possible electron-ion
collider~\cite{Gonderinger:2010yn}), further motivate revisiting this
observable. 

Although not necessarily linked to the problem of neutrino masses and
mixings (which signal the violation of neutral lepton flavours), cLFV
can also emerge in association with SM extensions incorporating a
mechanism of neutrino mass generation. 
Minimal extensions of the SM via additional sterile fermion states are
an appealing class of models, in particular those that succeed in explaining
oscillation data by the introduction of (not excessively) heavy states. 
Numerous studies have examined the impact of these models regarding 
several cLFV 
observables~\cite{Ma:1979px,Gronau:1984ct,Ilakovac:1994kj,Deppisch:2004fa,Deppisch:2005zm,Dinh:2012bp,Alonso:2012ji,Arganda:2014dta,Abada:2014kba,Abada:2014cca,Abada:2015zea,Abada:2015oba,DeRomeri:2016gum},
focusing either on specific realisations, or then evaluating the potential
contributions of sterile fermions via model-independent, simple
constructions (the so-called ``3+N'' models). 
 Among the many theoretically complete frameworks which simultaneously
explain neutrino data, while at the same time having a significant
phenomenological impact, one finds several low-scale seesaw
models, such as variants of a type I seesaw, the linear
seesaw~\cite{Akhmedov:1995vm,Malinsky:2005bi}, the Inverse Seesaw (ISS)
Seesaw~\cite{Mohapatra:1986bd} or the neutrino minimal SM
($\nu$MSM)~\cite{Asaka:2005an}.

In the present work we thus revisit cLFV in-flight conversions 
$\ell_i \to \ell_j$, carrying a full phenomenological analysis 
in the framework of SM extensions via sterile neutrinos. In
particular, we focus on flavour violating (FV) $Z$- and
photon-mediated interactions, recomputing their contributions, and
comparing our results to previous studies. We consider the three
different cLFV channels ($e - \mu$, $e - \tau$ and $\mu - \tau$), and
discuss the corresponding experimental prospects, confronting the
latter with other cLFV observables. 
In this study, we consider quasi-elastic scattering for the in-flight cLFV observables, 
which offers a first estimate due to the simple final state topology and to the associated background.
Although we do discuss the
potential of well-motivated low-scale seesaw models (in particular the
Inverse Seesaw and the $\nu$MSM), a first phenomenological approach -
and a significant part of the discussion - 
is done by means of an ``ad-hoc'' construction, a simple ``3+1 toy
model'', in which a single massive Majorana state is added to the SM
content, with no hypothesis on the underlying mechanism of mass
generation.

Our work is organised as follows: after describing the underlying
theoretical framework in Section~\ref{sec:SM+nus}, we discuss 
$\ell_i \to \ell_j$ in-flight conversions, including contributions to the
differential cross section and general features of the observables. 
The experimental prospects, as well as a comparative study with
other cLFV observables in minimal SM extensions via sterile neutrinos
are collected in Section~\ref{sec:res}; a brief overview as well as
further elements of discussion are summarised in the Conclusions.   
The relevant expressions of the $\gamma$- and $Z$-mediated interactions, 
together
with other relevant form factors, can be found in Appendices A and B.

\section{Minimal SM extensions via sterile fermions}\label{sec:SM+nus}

Motivated by several cosmological and experimental observations, 
sterile fermions are present as constituent blocks of many SM
extensions which encompass a mechanism of neutrino mass generation. 
If on the one hand sterile neutrinos can indeed provide an explanation
to the problem of neutrino masses and mixings, they can also open the
door to a rich phenomenology, with potential effects in a large
number of observables. This is a direct consequence of 
their mixings with the light (mostly active) neutrinos, which - if
non-negligible - lead to the violation of lepton flavour in both
neutral and charged leptonic currents~\cite{Schechter:1980gr,Gronau:1984ct}. 

In the presence of $n_S$ additional sterile (Majorana) neutrinos, the
vector and scalar currents are modified as
follows\footnote{Likewise, the interactions with
  neutral and charged Goldstone bosons are also modified:
 $\mathcal{L}_{G^0}\, =\,\frac{i g_w}{2 M_W} \, G^0 \,
\sum_{i,j=1}^{3 + n_S} {\bf C}_{ij}  \bar \nu_i  
\left(P_R m_j  - P_L m_i  \right) \nu_j\,+ \, \text{H.c.}$;   
$\mathcal{L}_{G^\pm}\, =\, -\frac{g_w}{\sqrt{2} M_W} \, G^- \,
\sum_{\alpha=1}^{3}\sum_{j=1}^{3 + n_S} {\bf U}_{\alpha j}   
\bar \ell_\alpha\left(
m_i P_L - m_j P_R \right) \nu_j\, + \, \text{H.c.}$.}  
(working in the physical basis, i.e., for mass eigenstates):
\begin{align}\label{eq:lagrangian:WGHZ}
& \mathcal{L}_{W^\pm}\, =\, -\frac{g_w}{\sqrt{2}} \, W^-_\mu \,
\sum_{\alpha=1}^{3} \sum_{j=1}^{3 + n_S} {\bf U}_{\alpha j} \bar \ell_\alpha 
\gamma^\mu P_L \nu_j \, + \, \text{H.c.}\,, \nonumber \\
& \mathcal{L}_{Z^0}\, 
= \,-\frac{g_w}{2 \cos \theta_w} \, Z_\mu \,
\sum_{i,j=1}^{3 + n_S} \bar \nu_i \gamma ^\mu \left(
P_L {\bf C}_{ij} - P_R {\bf C}_{ij}^* \right) \nu_j\,
-\frac{g_w}{4 \cos \theta_w} \, Z_\mu \,
\sum_{\alpha=1}^{3}  \bar \ell_\alpha \gamma ^\mu \left(
{\bf C}_{V} - {\bf C}_{A} \gamma_5 \right) \ell_\alpha\,, \nonumber \\
& \mathcal{L}_{H^0}\, = \, -\frac{g_w}{2 M_W} \, H^0  \,
\sum_{i,j=1}^{3 + n_S}  {\bf C}_{ij}  \bar \nu_i\left(
P_R m_i + P_L m_j \right) \nu_j + \, \text{H.c.}\,. 
\end{align}
In the above, $g_w$ denotes the weak coupling constant, 
$\cos^2 \theta_w = M_W^2 /M_Z^2$, 
$P_{L,R} = (1 \mp \gamma_5)/2$, 
and $m_i$ are the physical neutrino masses (light and heavy); 
the indices 
$\alpha $ denote the flavour of the charged leptons,
while $i, j = 1, \dots, 3+n_S$ correspond to the physical (massive) 
neutrino states. In addition 
${\bf C}_{V}$ and ${\bf C}_{A}$ are the SM coefficients
parametrizing the vector and axial-vector 
$Z$-couplings of charged leptons,   
${\bf C}_{V} = \frac{1}{2} + 2 \sin^2\theta_w$ and 
${\bf C}_{A} = \frac{1}{2}$.
Finally, a rectangular 
$3 \times (3 +n_S)$ mixing matrix, ${\bf U}_{\alpha j}$, parametrizes 
the mixing in charged current interactions (corresponding 
to the (unitary) PMNS matrix, $U_\text{PMNS}$ in the
case of $n_S=0$); the mixing between the left-handed leptons
corresponds to a $3 \times 3$ 
block of ${\bf  U}$, usually denoted $\tilde U_\text{PMNS}$.
The structure of ${\bf U}_{\alpha j}$ is at the source of 
lepton flavour violation in neutral currents, which, as seen from
above, is now parametrized by 
\begin{equation}
\label{eq:Cmatrix:def}
{\bf C}_{ij} \,=\,\sum_{\alpha=1}^{3} {\bf U}_{\alpha i}^*\,{\bf U}_{\alpha j}\,. 
\end{equation} 

\subsection{Constraints on sterile fermions}\label{sec:models:constraints}

Due  to the presence of the additional sterile states, 
the modified neutral and charged lepton currents 
might lead to new contributions to a
vast array of observables, possibly in conflict with current
data. These SM extensions via sterile fermions 
must be then confronted to all available constraints
arising from high-intensity, high-energy and cosmological
observations. 

In our subsequent phenomenological analysis, and for the  
theoretical framework considered, we ensure that compatibility with
the following constraints - theoretical (such as perturbativity of the
active-sterile couplings) and experimental - is verified at all times.

Sterile states, with a mass above the electroweak (EW) scale, can have
sizeable decay widths, a consequence of being sufficiently heavy to 
decay into a $W^\pm$ boson and a charged lepton, or into a 
light (active) neutrino and either a $Z$ or a Higgs boson.
One thus imposes the perturbative
unitarity
condition~\cite{Chanowitz:1978mv,Durand:1989zs,Korner:1992an,
Bernabeu:1993up,Fajfer:1998px,Ilakovac:1999md}, 
$\frac{\Gamma_{\nu_i}}{m_{\nu_i}}\, < \, \frac{1}{2}\, (i \geq 4)$. 
Noticing that the leading contribution to ${\Gamma_{\nu_i}}$
is due to the charged current term, one obtains the following 
bounds~\cite{Chanowitz:1978mv,Durand:1989zs,Korner:1992an,
Bernabeu:1993up,Fajfer:1998px,  
Ilakovac:1999md}:
\begin{equation}\label{eq:sterile:bounds:Ciimi}
m_{\nu_i}^2\,{\bf C}_{ii} \, < 2 \, \frac{M^2_W}{\alpha_w}\, \quad
\quad (i \geq 4)\,,
\end{equation}
where $\alpha_w=g^2_w/4 \pi$, and ${\bf C}_{ii}$ is given in
Eq.~(\ref{eq:Cmatrix:def}). 

Observational constraints on the sterile masses and their mixings with
the active states arise from an extensive number of sources. 
Firstly, and other than requiring compatibility between the left-handed lepton
mixing matrix $\tilde U_\text{PMNS}$ and the corresponding best-fit
intervals\footnote{We do not impose any constraints on the
(yet undetermined) value of the CP violating Dirac phase $\delta$.}
 defined from {\it $\nu$-oscillation
  data}~\cite{Tortola:2012te,Fogli:2012ua,GonzalezGarcia:2012sz,Forero:2014bxa,nufit,Gonzalez-Garcia:2014bfa,Esteban:2016qun},
we also impose, when relevant, {\it unitarity bounds} 
as arising from non-standard neutrino interactions with matter,
on the deviation 
of $\tilde U_\text{PMNS}$ from
unitarity~\cite{Antusch:2008tz,Antusch:2014woa,Blennow:2016jkn}.
Further constraints on the active-sterile
mixings (and on the mass regime of new states) arise from 
{\it electroweak precision observables}; these include 
new contributions to the invisible $Z$-decay width 
(addressed in~\cite{Akhmedov:2013hec,Basso:2013jka,
Fernandez-Martinez:2015hxa,Abada:2013aba}), which must comply 
with LEP results
on $\Gamma(Z \to \nu \nu)$~\cite{Olive:2016xmw}; 
moreover, any contribution to cLFV
$Z$ decay modes should not exceed the present uncertainty on the 
total $Z$ width~\cite{Olive:2016xmw}, $\Gamma (Z 
\to \ell_1^\mp \ell_2^\pm) < \delta \Gamma_{\rm  tot}$. In our study
we also take into account current limits on {\it invisible Higgs} decays 
(relevant for $m_{\nu_s} < M_H$), following the approach 
derived in~\cite{BhupalDev:2012zg,Cely:2012bz,Bandyopadhyay:2012px}. 
Likewise, negative results from 
{\it laboratory searches} for monochromatic lines in the
spectrum of muons from  $\pi^\pm \to \mu^\pm \nu$
decays are also taken into account~\cite{Kusenko:2009up,Atre:2009rg}.
As mentioned in the Introduction, the new states (through the modified
currents) induce potentially large contributions to cLFV observables; 
we evaluate the latter~\cite{Ma:1979px,Gronau:1984ct,Ilakovac:1994kj,Deppisch:2004fa,Deppisch:2005zm,Dinh:2012bp,Alonso:2012ji,Abada:2014kba,Abada:2015oba} 
imposing available limits on a wide variety of observables (some
of them collected in Table~\ref{table:cLFV:bounds}).  
In addition to the cLFV decays and transitions, which can
prove instrumental to test and 
disentangle these extensions of the SM, important constraints arise from
rare {\it leptonic and semileptonic decays of
pseudoscalar mesons} decays
(including lepton universality violating, cLFV and lepton number
violating modes); we include constraints from numerous 
$\ K,\ D$, $\ D_s$, $B$ modes (see~\cite{Goudzovski:2011tc,Lazzeroni:2012cx} for
kaon decays,~\cite{Naik:2009tk,Li:2011nij} for $D$ and $D_S$ decay
rates, and~\cite{Aubert:2007xj,Adachi:2012mm} for $B$-meson
observations), stressing that in the framework of the SM extended by
sterile neutrinos particularly severe constraints 
arise from the violation of lepton universality
in leptonic kaon decays (parametrized by the observable 
$\Delta r_K$)~\cite{Abada:2012mc,Abada:2013aba}. Finally, we also take
into account the recent constraints on 
{\it neutrinoless double beta decay}~\cite{Albert:2014awa}: should the sterile
states be Majorana fermions, they can potentially contribute to 
to the effective mass $m_{ee}$~\cite{Benes:2005hn}, which we evaluate 
following~\cite{Blennow:2010th,Abada:2014nwa}.

A number of {\it cosmological observations}~\cite{Smirnov:2006bu,Kusenko:2009up,Hernandez:2014fha,Vincent:2014rja}
put severe constraints on sterile neutrinos with a mass below the GeV
(in particular below 200~MeV). In our study we will in general explore
regimes associated with heavier sterile states ($m_{\nu_s} \gtrsim
0.5$~GeV) so that these constraints are not expected to play a relevant r\^ole.

\subsection{Theoretical framework}

Several mechanisms of neutrino mass generation, which in addition to
accommodating neutrino data, also address in the baryon asymmetry of
the Universe and/or put
forward a viable dark matter candidate, call upon sterile fermions. Among such
models, one encounters appealing SM extensions such as the Inverse 
Seesaw~\cite{Mohapatra:1986bd}, the  
$\nu$MSM~\cite{Asaka:2005an}, or several low-scale type I 
seesaw variants. 

\subsubsection{The simple ``3+1 model''}\label{sec:3+1}
As done in previous studies of cLFV in SM extensions via sterile 
neutrinos, one can use as a first phenomenological approach 
a minimal ``toy model'', consisting in the
addition of a single Majorana sterile neutral fermion to the SM field 
content~\cite{Abada:2014cca,Abada:2015oba}. 
 This ad-hoc construction makes no assumption on the mechanism
of neutrino mass generation; it thus allows to decouple 
 the neutrino mass generation (which could possibly
 arise at a different, higher 
scale, or stem from interactions not calling upon the lighter sterile
state) from the mechanism at the origin 
of flavour violation. In such a toy construction, 
the additional sterile state can also be
interpreted as encoding the effects of a larger number of states
possibly present in the model.

The simple toy model - which will be adopted in the present study
- thus relies on the simple hypothesis that the
interaction eigenstates and the physical ones are related via a
$4\times 4$ unitary mixing matrix, ${\bf U}_{ij}$. Other than the
masses of the three light (mostly active) neutrinos, and their mixing
parameters, the simple 
``3+1 model'' can be parametrized via the heavier (mostly sterile)
neutrino mass $m_4$, three active-sterile mixing angles as well as
three new CP violating phases (two Dirac and one Majorana). 
In the numerical analyses we will in general consider a
normal ordering for the light
neutrino spectra; in what concerns the new degrees of freedom, we 
will scan over the following range for the mass of the additional heavy
state, 
\begin{equation}\label{eq:effective:m4range}
0.5 \text{ GeV }\lesssim \, m_4 \, \lesssim 10^{6} \text{ GeV},
\end{equation}
while the active-sterile mixing angles are randomly taken to lie in the
interval $[0, 2 \pi]$ (as are the different CP violating phases).

\subsubsection{Complete theoretical frameworks for neutrino mass generation}
Several mechanisms of neutrino mass generation, which in addition to
accommodating neutrino data, address in addition the BAU and/or put
forward a viable DM candidate, call upon sterile fermions. 
Inverse seesaw realisations, as well as the $\nu$MSM, 
whose main features will be briefly 
summarised below, are an example of such extensions, known for their rich
phenomenological implications. \\

\noindent
{\it The (3,3) Inverse Seesaw realisation}

\noindent
The Inverse seesaw mechanism~\cite{Mohapatra:1986bd} 
relies in extending the SM via right
handed neutrinos and further sterile states. In the present analysis
we will consider a realisation of the ISS in which three generations
of RH neutrinos as well as three generations of extra singlet fermions
$X$ are added to the SM, $n_R = n_X = 3$; both $\nu_R$ and $X$ carry
lepton number, $L_R = L_X = +1$. The Lagrangian describing this
extension can be cast as 
\begin{equation}\label{eq:ISS:Ldef}
\mathcal{L}_\text{ISS} \,=\, 
\mathcal{L}_\text{SM} - Y^{\nu}_{ij}\, \bar{\nu}_{R i} \,\tilde{H}^\dagger  \,L_j 
- {M_R}_{ij} \, \bar{\nu}_{R i}\, X_j - 
\frac{1}{2} {\mu_X}_{ij} \,\bar{X}^c_i \,X_j + \, \text{H.c.}\,,
\end{equation}
with $\tilde{H} = i \sigma_2 H^*$ and $i,j = 1,2,3$ generation
indices. 
The light neutrino spectrum (containing mostly active states)
is given by a modified seesaw relation
\begin{equation}
m_{\nu} \, \approx \, \frac{(Y^\nu \, v)^2\, \mu_X}{M_R^2}
\end{equation}
where $\mu_X$ is the unique source of lepton number  
violation in the model. Small values of $\mu_X$ (which are thus
natural in the sense of 't Hooft) allow to accommodate 
the smallness of active neutrino masses for sizeable values of
$Y^\nu$, and hence a comparatively low seesaw scale 
($M_R$ lying close to the TeV scale). The spectrum of the (3,3) ISS
further contains three nearly degenerate pseudo-Dirac pairs; these
heavier, mostly sterile states have masses close to $M_R$ (their
degeneracy being lifted by $\mu_X$). 
The full $9\times 9$ mass matrix, $\mathcal{M}_\text{ISS}$, 
can be diagonalised as ${\bf U}^T \mathcal{M} {\bf U} 
= \text{diag}(m_i)$, with $i=1...9$. 
In the physical charged lepton basis, the leptonic mixings are encoded
in the rectangular sub-matrix ($3 \times 9$) 
defined by the first three columns of
${\bf U}$, its upper $3 \times 3$ block corresponding to the
non-unitarity $\tilde U_\text{PMNS}$. 

Depending on the specific realisation, and on the regimes for $M_R$
and $\mu_X$, the ISS can further account for the observed BAU via
leptogenesis~\cite{Abada:2015rta}, as well as provide viable 
DM candidates whose
relic density is in agreement with present observations, and which
could also accommodate possible indirect DM detection signals (if
confirmed)~\cite{Abada:2014vea,Abada:2014zra}.

\medskip

\noindent
{\it The $\nu$ Minimal Standard Model}

\noindent
The $\nu$MSM minimally extends the SM via the
inclusion of three RH neutrinos, aiming at simultaneously
addressing the problems of neutrino mass generation, 
the BAU and providing a viable DM
candidate~\cite{Asaka:2005an,Asaka:2005pn,Shaposhnikov:2008pf,
Canetti:2012kh}. The new particle content leads to new terms in the 
leptonic Lagrangian: 
\begin{equation}\label{eq:nuMSM:Lmass}
\mathcal{L}^\text{$\nu$MSM}_\text{mass} \, = \, 
-Y^\nu_{ij}\,\bar \nu_{Ri}\, \tilde H^\dagger L_j \, -\, 
\frac{1}{2}\, \bar \nu_{Ri}\, {M_{M}}_{ij}\,\nu^c_{Rj} + \text{H.c.}\,,
\end{equation}
where $i,j=1,2,3$ are generation indices, $L$ is the $SU(2)_L$
lepton doublet and $\tilde H = i \sigma_2 H^*$; $Y^\nu$ denotes the 
Yukawa couplings, while $M_{M}$ is a Majorana mass matrix (leading to
the violation of the total lepton number, $\Delta L=2$).

Other than three light (mostly active) neutrinos, the spectrum
contains three heavy states (with
masses $m_{\nu_{4-6}}$), whose masses and mixings to the lighter
states are strongly constrained in the case in which the $\nu$MSM is
called to successfully address the BAU and the DM problems.

\section{cLFV in-flight $\pmb{\ell_i \to \ell_j}$ conversion
}\label{sec:diff.cross}   
In what follows we summarise the most relevant points regarding the
computation of the observables associated with the in-flight cLFV
conversion; due to the underlying process, in which an intense lepton
beam hits a fixed target, the observable is also frequently referred to
as an ``on target'' cLFV transition, $\ell_i + N \to \ell_j +N^{(\prime)}$. 
As mentioned in the Introduction, there are several possibilities
regarding the final state of the nuclei (target) after interaction
with the energetic $\ell_i$ beam: elastic
scattering, in which $N=N^\prime$; quasi-elastic scattering, leading
to a final state target composed of several bodies (but conserving the
total number of nucleons, with no new hadronic states); 
inelastic processes (including excited nuclear states), and/or
nuclear fragmentation with associated pion or other light hadron
production (DIS regime). 
In the present phenomenological analysis we will focus on
the case of elastic scatterings\footnote{While inelastic scattering is expected to become dominant at large enough
$Q^2$, e.g. above $1~\text{GeV}^2$ for electron-proton scattering, its
description is beyond the purpose of this
paper.}; quasi-elastic processes (as well as
inelastic ones) were also recently 
addressed in the study of~\cite{Liao:2015hqu}). 

The kinematics of the in-flight cLFV conversion requires the beam to
have a minimal threshold energy (which depends on the nature of the
target and on the mass of final state lepton). Denoting the
intervening quadri-momenta as 
\begin{equation}\label{eq:def:kinematics}
\ell_i (k) \,+\, T (p)\,\to \,\ell_j (k^\prime) \,+\, T (p^\prime)\,, 
\quad \text{with} \quad Q^2\,=\, -q^2\,=\,-(k - k^\prime)^2\,=\,
2\, M_T\, \Delta E_\text{beam}\,,
\end{equation}
with $\Delta E_\text{beam}$ the energy loss of the beam, and 
$M_T$ the target's mass, one thus finds that the (threshold) beam
energy is\footnote{While Eq.~(\ref{eq:beam:threshold}) leads to an 
  effective lower bound to the beam energy,  as previously mentioned 
  we will not enter high-energy regimes leading to DIS phenomena.} 
\begin{equation}\label{eq:beam:threshold} 
E_\text{beam}\,  > \, m_{\ell_j} \, \left ( 1 + \frac{m_{\ell_j}}{2
  M_T} \right)\,,
\end{equation}
in which $m_{\ell_j}$ denotes the mass of the heavier lepton in the final
state (muon or tau). Moreover, a non-zero
momentum transfer to the nuclear system is unavoidable. Depending on
the beam's energy, and the composition of the target, one finds
minimal values for the energy transfer - although these do decrease
with increasing beam energy and with the (larger) size  of the 
nuclei, non-zero values of $Q^2$ are
always obtained (see~\cite{Liao:2015hqu} for a comprehensive
discussion). 

In the framework of NP models in which cLFV occurs via
higher-order (loop) transitions (as is the case of R-parity conserving
SUSY, seesaw realisations, etc.), 
the differential cross section for the cLFV conversion of
Eq.~(\ref{eq:def:kinematics}) receives contributions from different 
processes, depending on the interaction(s) at the source of flavour violation: 
 photon dipole, $Z$- and Higgs-penguins, box diagrams, among other contributions.
In what follows, we proceed to discuss them.  

\bigskip
The differential cross section for the on-target conversion of 
$\ell_i \to \ell_j$, 
exclusively due to photon dipole exchanges (i.e., putting to zero all other
contributions), can be written as~\cite{Liao:2015hqu}
\begin{equation}\label{eq:photon:dsigma.dQ2:photon}
\left. \frac{d \sigma^{i\to j}}{d Q^2} \right|_{\gamma}\, =\, 
\frac{\pi \, Z^2\, \alpha^2}{Q^4 \, E_\text{beam}^2}\, 
H^\gamma_{\mu \nu}\, L^{\gamma \mu \nu}_{ij}\,,
\end{equation}
in which $Z$ denotes the target atomic number. 
The detailed expression for the hadronic
tensor $H^\gamma_{\mu \nu}$ can be found in the
Appendix~\ref{appendix:dsigma:computation}, while the leptonic tensor
can be decomposed as 
\begin{equation}\label{eq:photon:dsigma.dQ2:photon:Lmunu}
L^{\gamma \mu \nu}_{ij}\, =\, L^\gamma_{ij}\, L^{\gamma \mu \nu}(k, q)\, ,
\end{equation}
in which $L^\gamma_{ij}$ encodes the cLFV (effective) couplings.
Important contributions to the on-target cLFV conversion arise from the
$Z$-mediated interaction. Likewise, and in 
the limiting case in which only $Z$-interactions are present, 
one can write 
\begin{equation}\label{eq:photon:dsigma.dQ2:Z}
\left. \frac{d \sigma^{i\to j}}{d Q^2} \right|_{Z}\, =\, 
\frac{G_F^2}{32 \,\pi\, E_\text{beam}^2}\, 
H^Z_{\mu \nu}\, L^{Z \mu \nu}_{ij}\,,
\end{equation}
with 
\begin{equation}\label{eq:photon:dsigma.dQ2:Z:Lmunu}
L^{Z \mu \nu}_{ij}\, =\,  L^{Z}_{ij}\, L^{Z \mu \nu}(k, q)\, ,
\end{equation}
where, and as before, the terms $L^Z_{ij}$ encode the cLFV couplings.
Other contributions, such as Higgs mediated interactions (as in the
case of SUSY models), box diagrams, etc., might be also present and,
depending on the given model (and regime), play a relevant r\^ole. 

\begin{center}
\begin{figure}
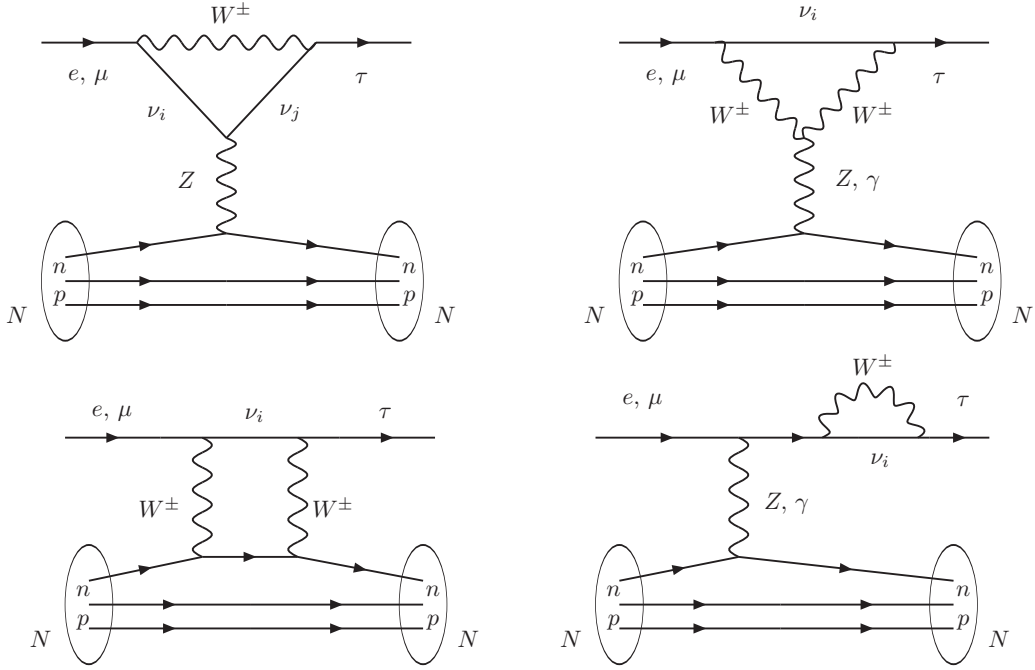

\begin{center}
\begin{tabular}{cc}
\epsfig{file=figs/mu-tau.penguin.1nu.epsi, width=60mm}
\hspace*{5mm}&\hspace*{5mm}
\epsfig{file=figs/mu-tau.penguin.2nu.epsi, width=60mm}
\\
\epsfig{file=figs/mu-tau.penguin.3nu.epsi, width=60mm}
&
\epsfig{file=figs/mu-tau.penguin.4nu.epsi, width=60mm}
\end{tabular}
\end{center}
\caption{Contributions to the 
$\ell_i - \tau$ conversion from $Z$- and $\gamma$-penguins,  
``handbag'' (box diagrams) and cLFV corrections to the lepton propagator. }
\label{fig:diagrams}
\end{figure}
\end{center}

The $L^{\gamma,Z}_{ij}$ couplings can be interpreted
as generic sources of flavour violation 
at the origin of the cLFV in-flight conversion, in
the framework of SM extensions in which cLFV receives important (if
not dominant) contributions from penguin loop diagrams; however, and 
in what
follows we focus on a minimal NP model: the SM minimally extended by
additional  (massive) neutrinos. In such a framework, 
the most important contributions indeed arise 
from $Z$ and photon mediated interactions,
$W^\pm$ mediated box diagrams, and corrections to the lepton
propagators, some of them schematically depicted in
Fig.~\ref{fig:diagrams}. 
In our analysis, we will not take into account the contributions
arising from the ``handbag'' (box) diagrams as in
the limiting (unrealistic) 
case of a real quark, these diagrams would
correspond to the usual box contributions common to several
observables (such as $\mu-e$ conversion in Nuclei, or $\mu \to 3 e$
decays). In minimal SM extensions via sterile fermions, and in 
the large sterile mass regime - which has been shown to be
associated with sizeable contributions to the above mentioned decays - 
the box contributions typically lead to  
subdominant contributions when compared to the $\gamma$
and $Z$
penguins~\cite{Abada:2014kba,Abada:2014cca,Abada:2015oba,Abada:2015zea}. Other
regimes are known to be associated with important box-diagram
contributions~\cite{Abada:2014kba}. 
It is worth stressing that the Wilson coefficients
 for the contribution of boxes, photon and $Z$-penguins
 (cf. Fig.~\ref{fig:diagrams}) have been evaluated 
 in the SM extended by sterile massive fermions
 with non-negligible active-sterile mixings~\cite{Abada:2015zea}, 
 with results confirming the above
  statement.

In this context, 
$L^\gamma_{ij}$ can be cast as 
\begin{equation}\label{eq:photon:dsigma.dQ2:photon:Lij}
L^\gamma_{ij}\, =\, \frac{\alpha_w^3\, s_w^2}{64\, \pi \,e^2}\,
\frac{m_{\ell_j}^2}{M_W^4}\, \left| G^\gamma_{ji}\right|^2\,,
\end{equation}
with $G^\gamma_{ji}$ denoting the photon-lepton dipole coupling, 
also contributing to other cLFV
transitions such as $\ell_j \to \ell_i \gamma$, and which is given in
Appendix~\ref{appendix:cLFV:formulae};
the flavour violating $Z$-couplings can be written
\begin{equation}\label{eq:photon:dsigma.dQ2:Z:Lijpm}
L^Z_{ij}\, =\, \frac{\alpha_w^4}{G_F^2\, M_W^4} 
\frac{2(-1/2 + \sin^2_w)^2 + \sin^4_w}{64}\, \left| F^Z_{ji}\right|^2\, ,
\end{equation}
in which $F^Z_{ji}$ denotes the form factor encoding flavour violating 
$Z \ell_j \ell_i$ interactions, which is also present in several
other cLFV observables (see Appendix~\ref{appendix:cLFV:formulae}).  
The full expressions for $d \sigma^{i\to j}/{d Q^2}|_{\gamma, Z}$,
as well as that of full leptonic and hadronic tensors are given in 
Appendix~\ref{appendix:dsigma:computation}.

While in low sterile mass regimes the photon penguin does dominate
over the $Z$, 
increasing the mass of the sterile neutrinos - which corresponds to
regimes typically associated with a significant enhancement of 
the contributions to many cLFV (in particular to 
the in-flight differential cross sections under study) -  
leads to having a $Z$-penguin contribution which
increasingly dominates over the photon-ones. 
Although this cannot be straightforwardly inferred by comparing 
Eq.~(\ref{eq:photon:dsigma.dQ2:photon}) and
  Eq.~(\ref{eq:photon:dsigma.dQ2:Z}) 
- since the source of cLFV is encoded in
the form factors $G^{\gamma}_{ij}$ and $F^{Z}_{ij}$ of
of  Eqs. (\ref{eq:photon:dsigma.dQ2:photon:Lij},
  \ref{eq:photon:dsigma.dQ2:Z:Lijpm}), respectively
- we notice that   
contrary to diagrams in
which a single neutrino and a $W^\pm$ run in the loop
(see Fig.~\ref{fig:diagrams}, upper-right diagrams), the $Z$-penguin
further receives contributions from loops where two neutrino states 
and one $W$ boson are present (Fig.~\ref{fig:diagrams}, upper-left
diagram).

As is clear from the above discussion concerning the cLFV couplings, 
current bounds on many low-energy observables (see
Table~\ref{table:cLFV:bounds})  
will play a very constraining r\^ole on the
maximal viable values for the in-flight conversion 
cross section. Particularly relevant will prove to be
the bounds from $\ell_j \to 3\ell_i$ decays, radiative decays, as well
as $\mu-e$ conversion in Nuclei. 

\begin{figure}
\begin{center}
\begin{tabular}{cc}
\hspace*{-7mm}
\epsfig{file=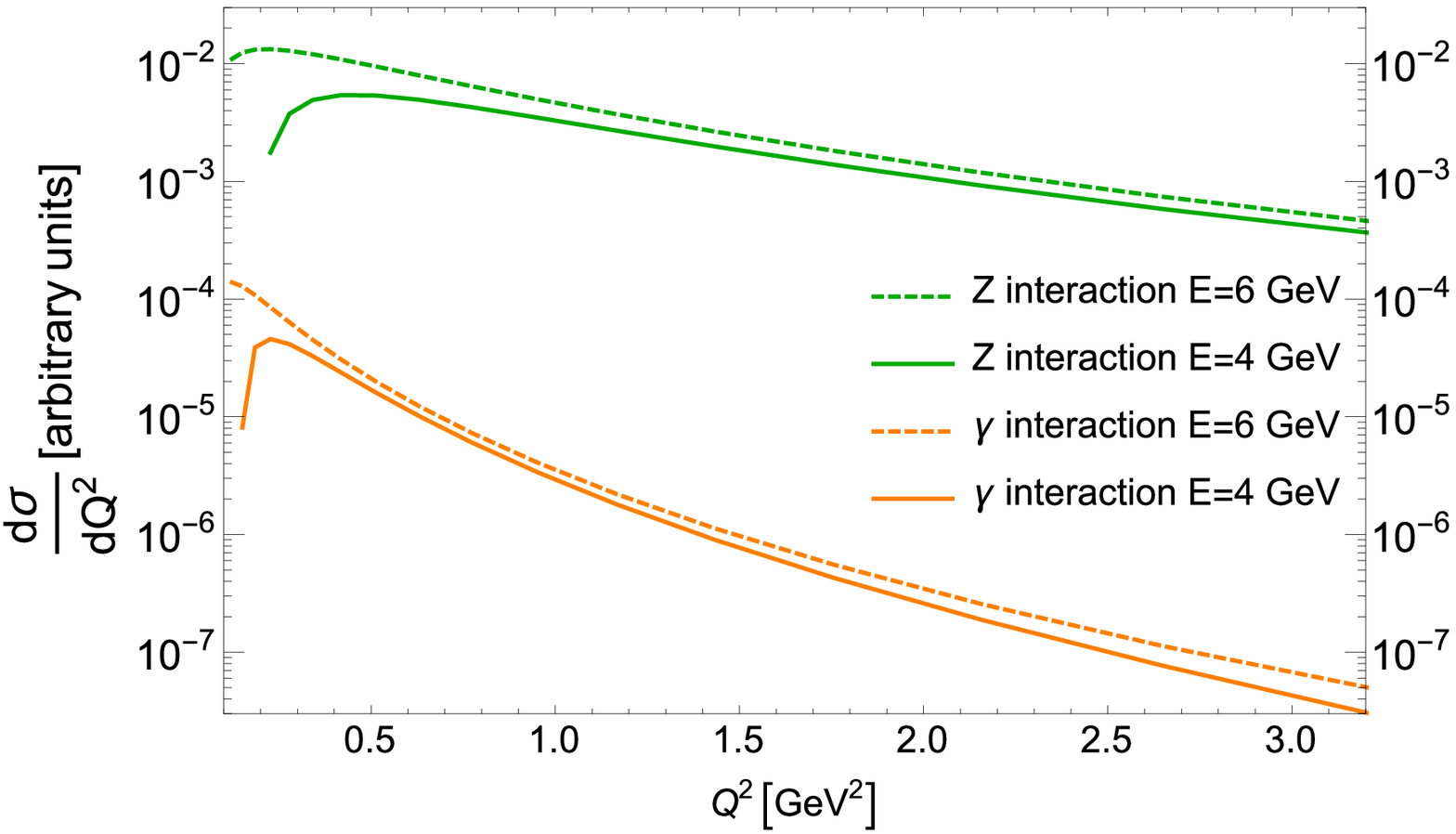, width=80mm } 
\hspace*{1mm}&\hspace*{1mm}
\epsfig{file=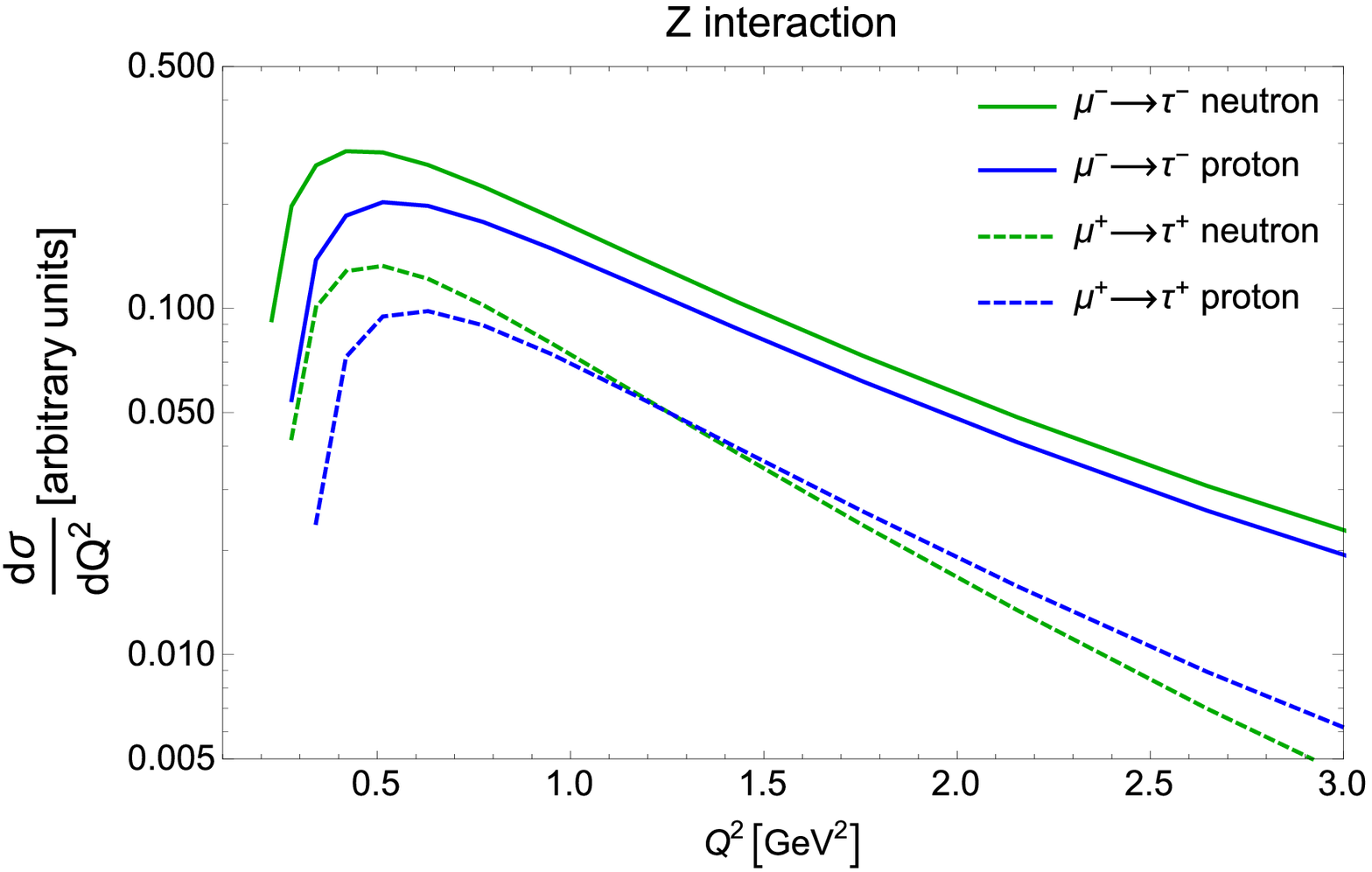, width=78mm }
\end{tabular}
\end{center}
\caption{
Differential cross section (arbitrary units) for nucleon targets  
as a function of the momentum transfer, $Q^2$. 
On the left panel, comparison of the individual $Z$ (green) and photon
(orange) interactions, for $\mu-\tau$ conversion on a neutron target;
full (dashed) lines correspond to $E=4\,(6)$~GeV beam energy.
On the right panel, $Z$-mediated $\mu-\tau$ conversion
for neutron (green) and proton (blue) targets, with 
full (dashed) lines correspond to lepton (anti-lepton) conversion.
}\label{figs:diffcross.Q2} 
\end{figure}

\bigskip
Before entering the study of the prospects for the cLFV on-target
conversion
in extensions of the SM via sterile fermions, we briefly discuss some
issues regarding the nuclear interaction and the beam energy, which can
be already understood from the differential cross section, 
$d \sigma^{i \to j}/ d Q^2$. 
The nuclear tensors - for both photon and $Z$-mediated interactions - 
can be computed for either spin 0 and spin 1/2 targets. In our
phenomenological study, we consider elastic interactions with individual
nucleons, that is with spin 1/2 protons and/or neutrons
(which corresponds to setting $M_T = M_{p, n}$ and $Z=1$ 
in the relevant equations). 
The individual differential cross sections, corresponding to the purely
$Z$- or $\gamma$-mediated  exchanges, for $\mu-\tau$ conversion on a neutron
target, are displayed on the left panel of 
Fig.~\ref{figs:diffcross.Q2} as a function of the momentum transfer,
$Q^2$, and for two different beam energies, $E=4\,,6$~GeV. 
These have been evaluated
by simply setting by hand, in a model-blind manner,
maximal values for the flavour violating terms $L^{Z,\gamma}_{ij}$,   
see Eqs.~(\ref{eq:photon:dsigma.dQ2:photon:Lij}, 
\ref{eq:photon:dsigma.dQ2:Z:Lijpm}). (Leading to the results displayed
in this section, no observational bounds have been applied.)
Although depending on the actual SM extension under consideration (and
in the specific case of additional sterile fermions, on the particular
mass regime), $Z$-mediated FV conversions often prove to dominate over
the  photon dipole exchanges (see~\cite{Abada:2014kba,Abada:2014cca,Abada:2015oba}), the example seen in the left panel of
  Fig.~\ref{figs:diffcross.Q2} 
 being typical of heavy sterile masses in the 1-10~TeV range.

Unless otherwise stated, in the
following numerical discussion, we will in general consider that $Z$-penguins
provide the dominant contributions to the observables under study.  

On the right panel of Fig.~\ref{figs:diffcross.Q2} we compare the 
$Z$-mediated contribution for the individual nucleons (proton and neutron).
In view of the very similar behaviour for both nucleons, in the
following we will for simplicity assume a neutron target (unless
otherwise explicitly mentioned). Likewise, and in agreement with the
findings of~\cite{Liao:2015hqu}, there is only a small difference,
typically below $40\%$, 
regarding the differential cross section associated with the 
cLFV conversion of leptons or anti-leptons (cf.
Eq.~(\ref{eq:photon:dsigma.dQ2:Z:Hmunu}),
Appendix~\ref{appendix:dsigma:computation}); thus in our analysis we
will discuss $\ell^{-}_i n \to  \ell^-_j n $. 
Even though the results displayed in Fig.~\ref{figs:diffcross.Q2}
correspond to $\mu-\tau$ conversion, qualitatively analogous ones have
been found for an electron beam (with final state muons or taus). 

\begin{figure}
\begin{center}
\begin{tabular}{cc}
\hspace*{-7mm}
\epsfig{file=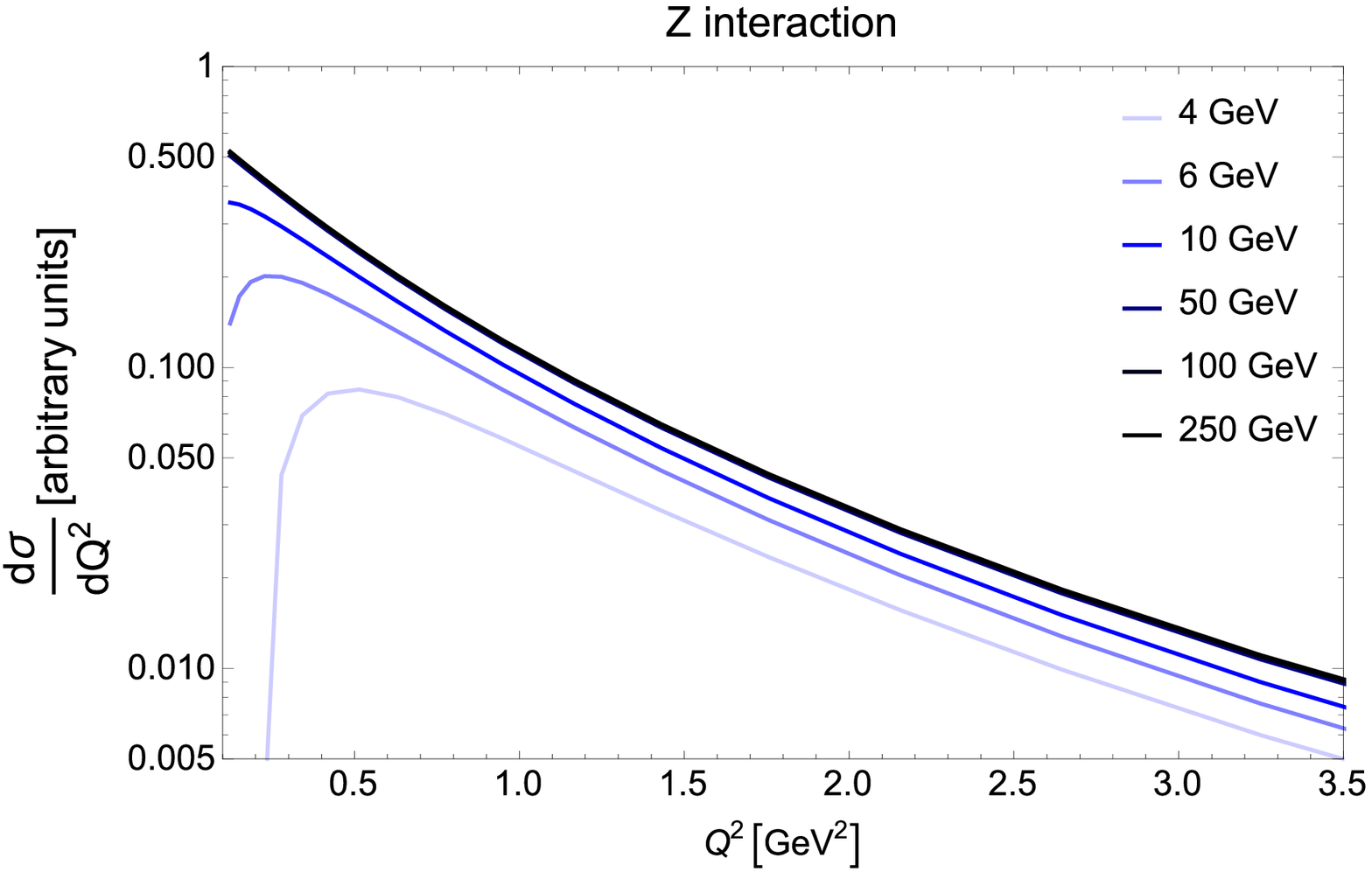, width=80mm}
\hspace*{1mm}&\hspace*{1mm}
\epsfig{file=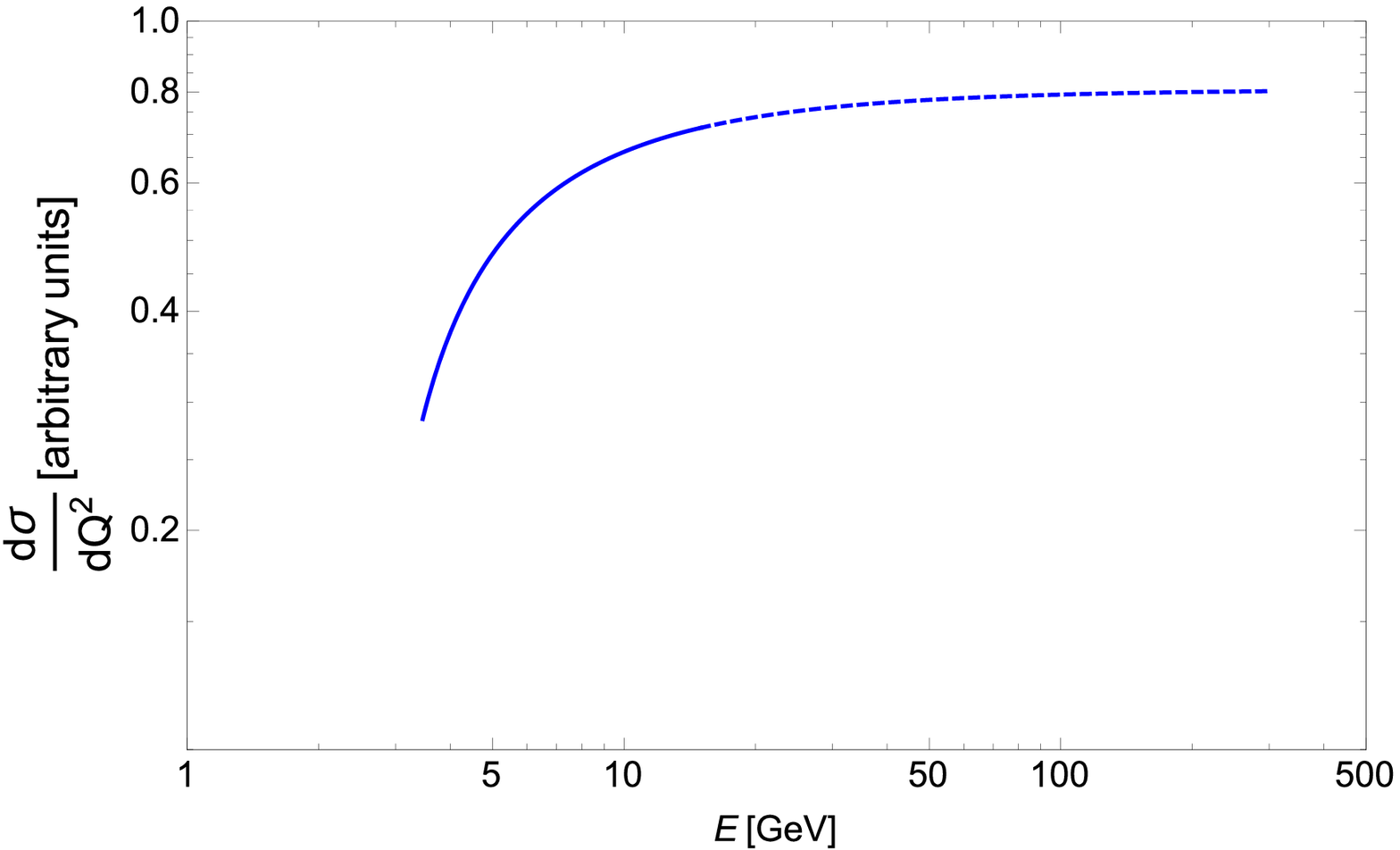, width=77mm }
\end{tabular}
\end{center}
\caption{On the left, 
differential cross section (arbitrary units)   
as a function of the momentum transfer, $Q^2$, for different beam energies. 
On the right, variation of the differential cross section for
$\mu-\tau$ conversion on a neutron target (arbitrary
units) with the beam energy for a fixed value of $Q^2$; 
full (dashed) lines denote the elastic scattering (na\"ive extrapolation
to DIS regime).
}\label{figs:diffcross.Ebeam} 
\end{figure}

A second comment concerns
the dependency of the differential cross section on 
the beam's energy, which was already manifest in the results of
Fig.~\ref{figs:diffcross.Q2}.  
Although both photon and $Z$ mediated contributions 
explicitly scale as $E^2_\text{beam}$, the hadronic tensors (see
Appendix~\ref{appendix:dsigma:computation}) 
both have non-trivial dependencies (also via $Q^2$ -
cf.~Eq.~(\ref{eq:def:kinematics})). The left panel of
Fig.~\ref{figs:diffcross.Ebeam} generalises the choices of beam energy,
$E=4\,(6)$~GeV, presented in Fig.~\ref{figs:diffcross.Q2}; 
for larger values of the beam energy one enters the strong DIS regime
- in the latter
case, the behaviour of the differential cross section must be
interpreted as only illustrative 
(the results here computed no longer quantitatively hold). For a fixed
value of the momentum transfer $Q^2$ (which maximises
the conversion rate), the dependency of the
differential cross section on the beam energy 
is illustrated on the right panel 
of Fig.~\ref{figs:diffcross.Ebeam}. The latter
confirms that once the beam energy is sufficiently large to
reach the threshold for the in-flight conversion to occur, see
Eq.~(\ref{eq:beam:threshold}), the rate mildly increases until
rapidly saturating (in the displayed case at $E_\text{beam} \approx
10$~GeV).
 
\section{Experimental prospects}\label{sec:res}

The total expected number of  produced leptons $\ell_j$  
for the in-flight $\ell_i \to
\ell_j$ conversion can be written as  
\begin{equation} 
N_\text{conver.}(\ell_i \to \ell_j)\, =\, 
N_{\ell_i}\, \times \, P(\ell_i \to \ell_j)\,,
\end{equation} 
where $N_{\ell_i}$ denotes the number of leptons ($e, \,\mu$) hitting
the target, and $P(\ell_i \to \ell_j)$ the conversion probability. 
For the case of $e \to \mu$ conversion, the total 
number of signal events can be directly obtained from the above
equation, simply rescaling $N_\text{conver.}(e \to \mu)$ via
parameters associated with the specificity of the target (thickness 
$L$ and density $\rho$, or equivalently, 
the target's mass $T_m$ - expressed in g/cm$^2$). 
For the case of final state tau leptons, their
average lifetime implies that they will rapidly decay, and hence one has a
further correction factor of BR($\tau \to \mu \nu \nu$), 
which in the SM is approximately 17.4\%~\cite{Olive:2016xmw}. 
Thus, the final number of expected conversions can be cast
as~\cite{Gninenko:2001id}   
\begin{equation}\label{eq:nevents}
N_\text{signal}(\ell_i \to \ell_j)\, =\, 
N_{\ell_i}\, \times \, \sigma(\ell_i \to \ell_j)\,
\times \, T_m \, \times \, N_{p+n}\,  
\left[\times \,\text{BR}(\tau \to \mu \nu \nu)\right]\,,
\end{equation} 
with $\sigma(\ell_i \to \ell_j)$ the integrated cross
section and $N_{p+n}$ the total number of nucleons per gramme of target -
assuming for simplicity an average value of the
contributions from protons and
neutrons to the total cLFV conversion cross section. One thus finds
\begin{equation}\label{eq:nevents:final}
N_\text{signal}(\ell_i \to \ell_j)\, =\, 
N_{\ell_i}\, \times \, \left( \frac{\sigma(\ell_i \to
  \ell_j)}{\text{fb}} \right) \,
\times \, \left( \frac{T_m}{\text{g cm}^{-2}} \right) 
\, \times \, 6 \times 10^{-16}\,  
\left[\times \,\text{BR}(\tau \to \mu \nu \nu)\right]\,.
\end{equation} 
Recall that in the above two equations, 
the last term $\text{BR}(\tau \to \mu \nu \nu)$ 
is only present when the final lepton is a $\tau$.
In order to discuss the real expected number of events, 
one should further take into consideration the detector's intrinsic 
efficiency,  
$\epsilon_\text{d}$, as well as the relevant contributions to the  
background - which we will not address in the present study. 

In Table~\ref{table:target.benchmarks} we collect some operating
benchmark values (surface density of the target and intensity of the
beam), previously considered in former discussions of this
cLFV observable.
\begin{table}[h!]
\centering
\begin{tabular}{|c|c|c|c|}
\hline
Facility & 
Beam nature  & $T_m$ & Intensity (leptons/yr) \\
\hline
Linear Collider & $e^\pm$ & 10 g/cm$^2$ & $10^{22}$ \\
Muon Collider ($\nu$-Factory) & $\mu^\pm$ 
& 100 g/cm$^2$ & $10^{20}$ \\
COMET & $\mu^-$ & $\sim$ 1 g/cm$^2$ (Al) &  $10^{19}$ \\
NA64 & $\mu^-$ & $\sim$ 1000 g/cm$^2$ (active) &  $10^{14 - 15}$\\
\hline
\end{tabular}
\caption{Illustrative benchmark values for 
surface density of target (in g/cm$^2$) as well as nature and
intensity of the potential beams used for in-flight cLFV conversion
(cf.~\cite{Cui:2009zz,Kanemura:2004jt,SPS:NA64,Gninenko:private}).}
\label{table:target.benchmarks}
\end{table}

\bigskip
\noindent
{\it The simple ``3+1 model''}

Hereafter focusing on the most minimal ``3+1 model'', 
described in Section~\ref{sec:3+1}, we begin our discussion of the
integrated cross section for the several cLFV in-flight conversion
modes; as an illustrative case, we
present the results obtained for a lepton beam energy of 4~GeV (independent
of its nature, electron or muon). Prospects for different (higher)
energy beams have already been briefly commented in the
previous section, and the qualitative outcome holds for the present discussion.
Moreover, and although having carried the numerical computation of
both $Z$ and photon penguin contributions, we only present the contributions
of the former, which in our framework 
are dominant with respect to those of the latter. 

\begin{figure}
\begin{center}
\begin{tabular}{cc}
\hspace*{-7mm}
\epsfig{file=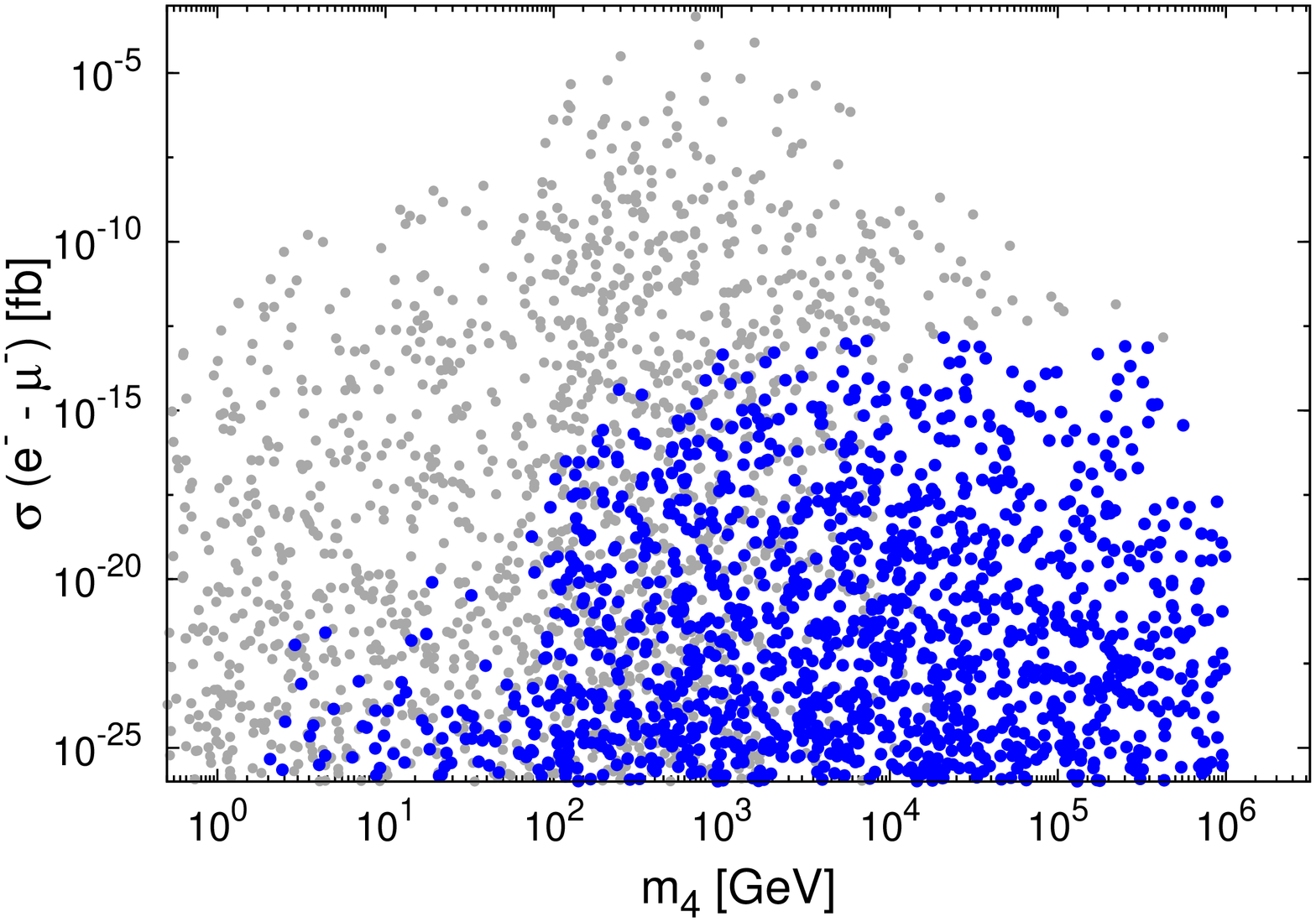,
  width=58mm , angle=-90 } 
& 
\epsfig{file=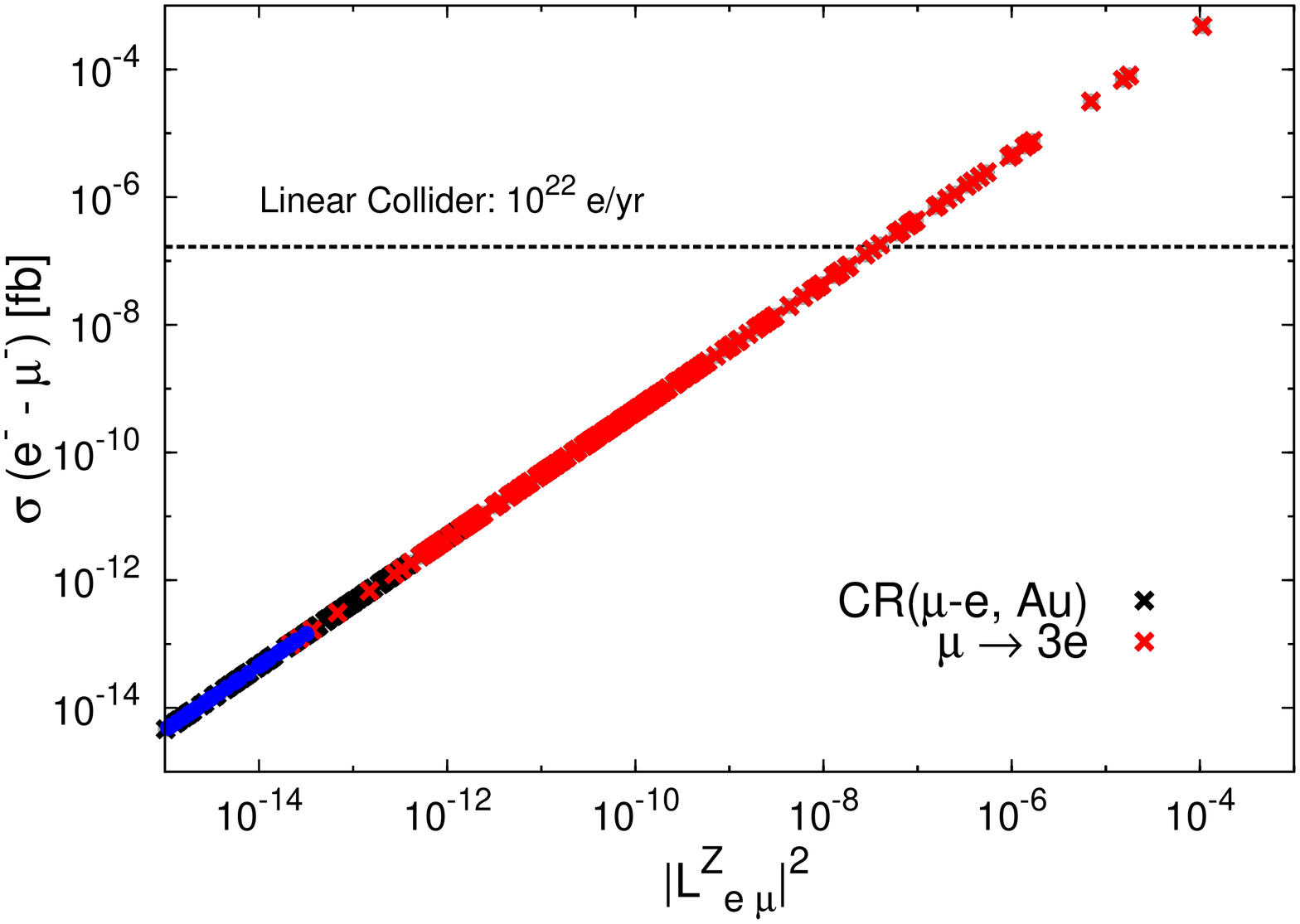,
  width=58mm , angle=-90 } 
\\ 
\hspace*{-7mm}
\epsfig{file=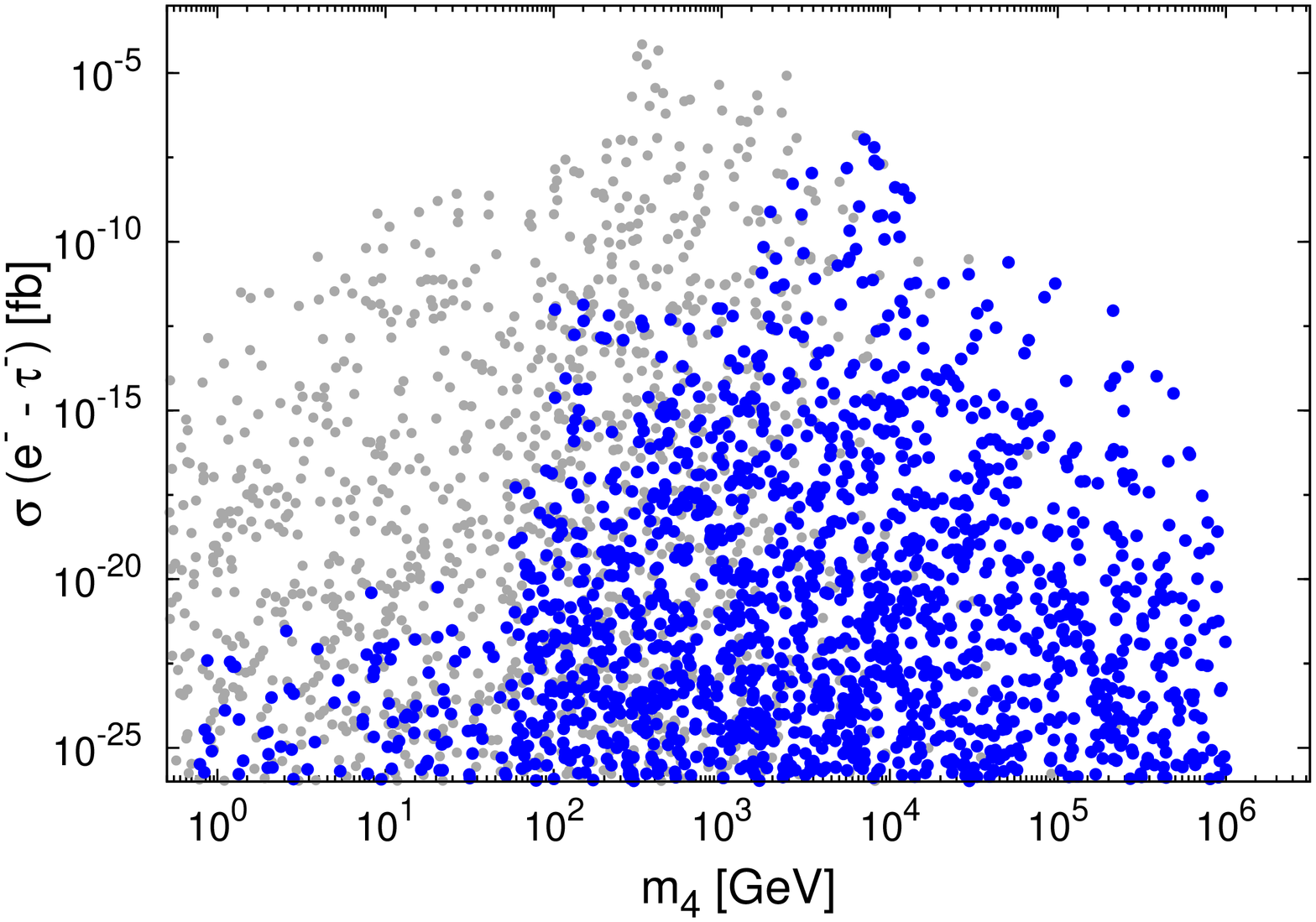,
  width=58mm, angle=-90 }
&
\epsfig{file=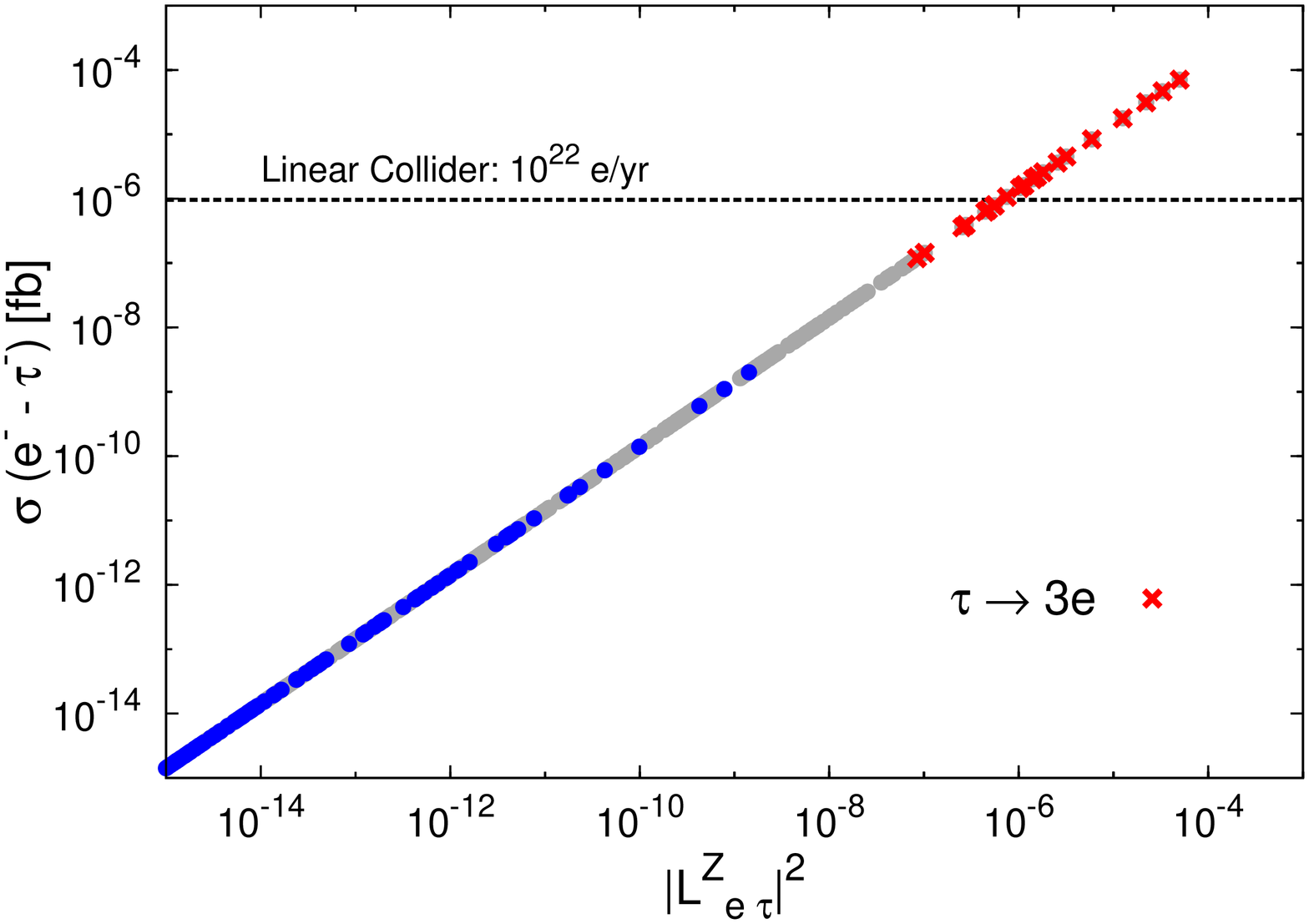,
  width=58mm , angle=-90 } 
\\
\hspace*{-7mm}
\epsfig{file=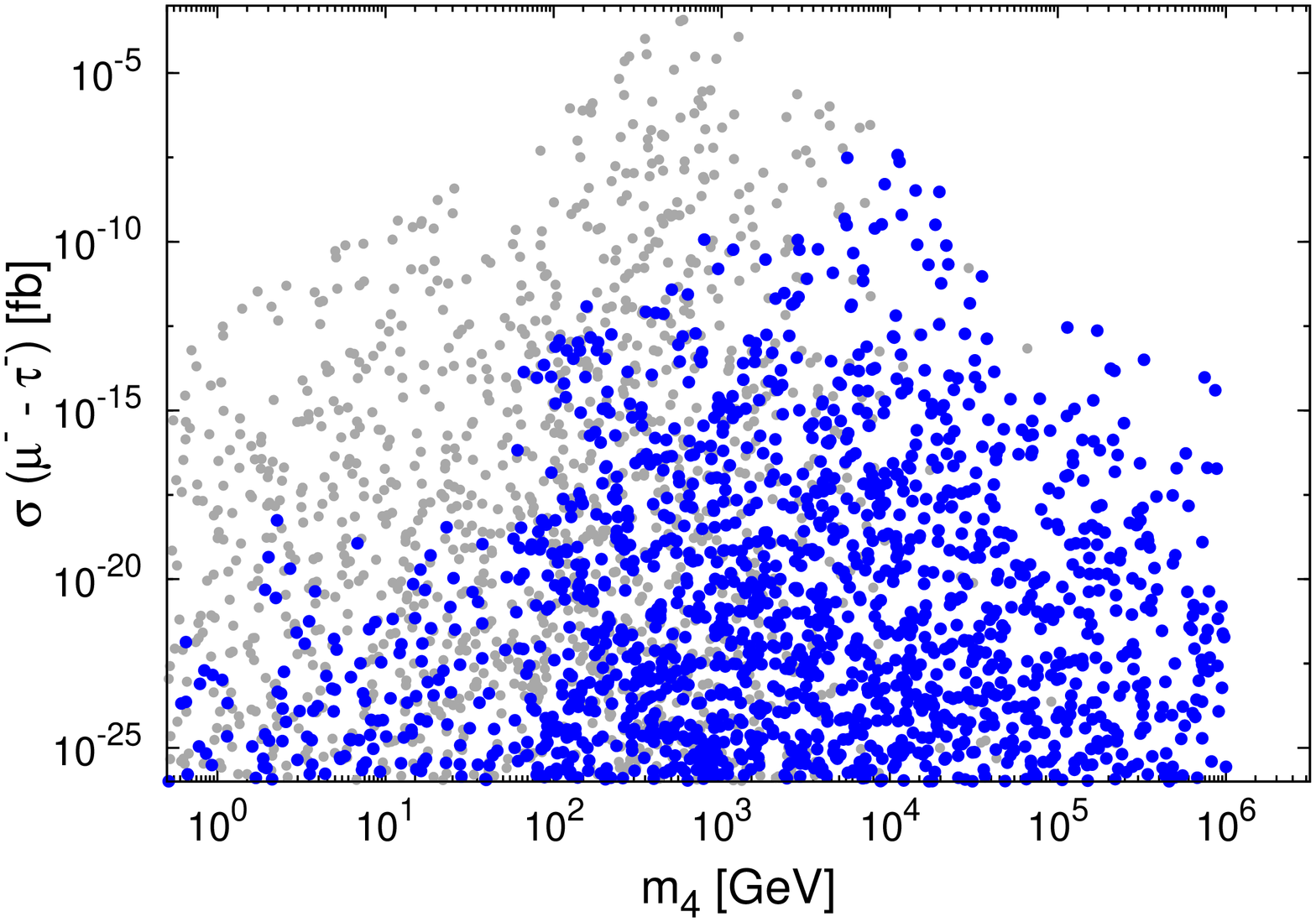,
  width=58mm, angle=-90 }
&
\epsfig{file=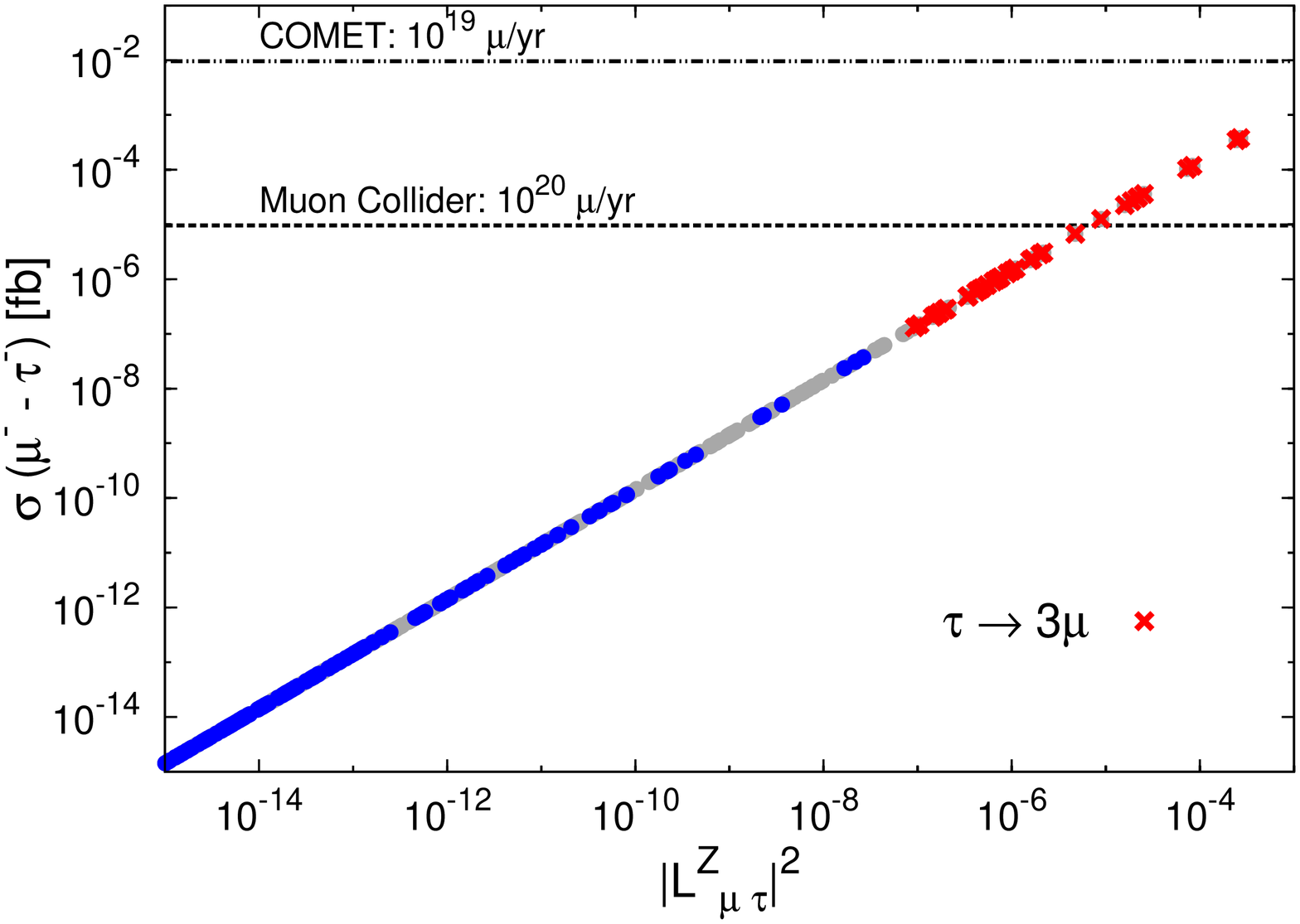,
  width=58mm , angle=-90 } 
\end{tabular}\vspace*{2mm}
\end{center}
\caption{``3+1 model": on the left, values of 
  $\sigma (\ell_i \to \ell_i )$ (in fb) 
  as a function of $m_4$ (in GeV), for a
  beam energy $E=4$~GeV. From top to bottom, $e \to \mu$, 
  $e \to \tau$ and $\mu \to \tau$. 
  Blue coloured points comply with the different constraints
  discussed in Section~\ref{sec:models:constraints}; those in grey
  violate at least one phenomenological and/or experimental bound.
  On the right column, $\sigma (\ell_i \to \ell_i )$ vs. the
  amount of flavour violation in the $Z$-mediated interaction,
  $|L^Z_{ij}|^2$ - see Eq.~(\ref{eq:photon:dsigma.dQ2:Z:Lijpm}); the additional
  colour coding of the points reflects the most stringent cLFV
  constraints in each case. Horizontal lines further denote 
  the cross sections leading to ``observable'' in-flight conversions
  for the appropriate benchmarks of
  Table~\ref{table:target.benchmarks}.  
  }\label{figs:sigma:eff.m4.cL} 
\end{figure}

The different panels of Fig.~\ref{figs:sigma:eff.m4.cL} display 
a general survey of the expected contributions 
to the different cross sections (arising from $Z$-mediated 
cLFV interactions), $\sigma (\ell_i \to \ell_j )$ 
as a function of the mass of the heavy, mostly sterile state. 
The left column of Fig.~\ref{figs:sigma:eff.m4.cL} confirms 
that the cLFV cross section rapidly increases for heavy neutrino masses
above the EW scale. Although one could potentially have values for the
different observables as large as $\sigma (\ell_i\to \ell_j) 
\approx \mathcal{O}(10^{-3})$, current experimental bounds -
in particular those arising from the violation of several cLFV bounds -
exclude these regimes. In terms of expected number of converted
leptons, having at least 10 conversions per year lies beyond 
realistic prospects for beam intensities: even for the least constrained
observable, $\mu \to \tau$ conversion, very intense muon beams on a dense target
cannot account for more than 0.04 converted tau leptons
per year (for $e \to \tau$ one would have at best 0.02 converted
$\tau$s, and even lower numbers for $e \to \mu$ conversion).

For final state tau leptons, the strongest cLFV
constraint arises from the corresponding 3-body decays ($\tau \to 3
\ell_i$), while for $e \to \mu$ conversion  
the current bounds on CR($\mu-e$, Au) further add to the 
already constraining r\^ole of $\mu \to 3e$.
The right hand side column of Fig.~\ref{figs:sigma:eff.m4.cL} 
summarises this discussion, displaying $\sigma (\ell_i \to \ell_i)$ as
a function of the flavour violation in $Z$-mediated interactions,
$|L^Z_{ij}|^2$ - see Eq.~(\ref{eq:photon:dsigma.dQ2:Z:Lijpm}), and
Appendix~\ref{appendix:dsigma:computation}. Horizontal lines
denote the cross sections that would account for a minimum of  
10 conversions per year 
(the different line scheme corresponding to the relevant operating 
benchmarks of Table~\ref{table:target.benchmarks}). 
Other than the coloured points associated with the leading cLFV
constraints, grey points are associated with further exclusions
arising from many other observables - as described in
Section~\ref{sec:models:constraints}. 

\begin{figure}
\begin{center}
\begin{tabular}{cc}
\epsfig{file=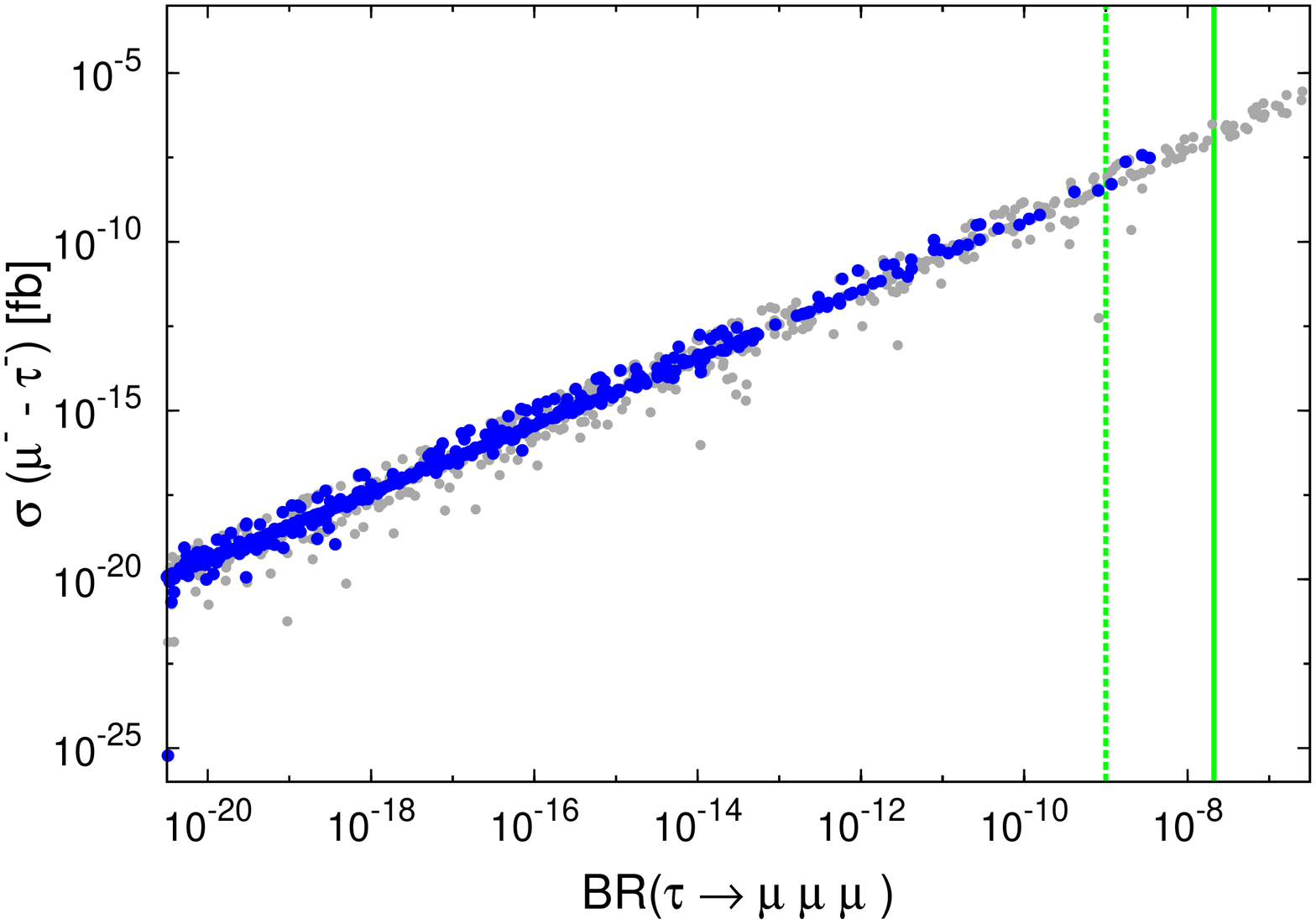,
  width=60mm, angle=-90 }
&
\epsfig{file=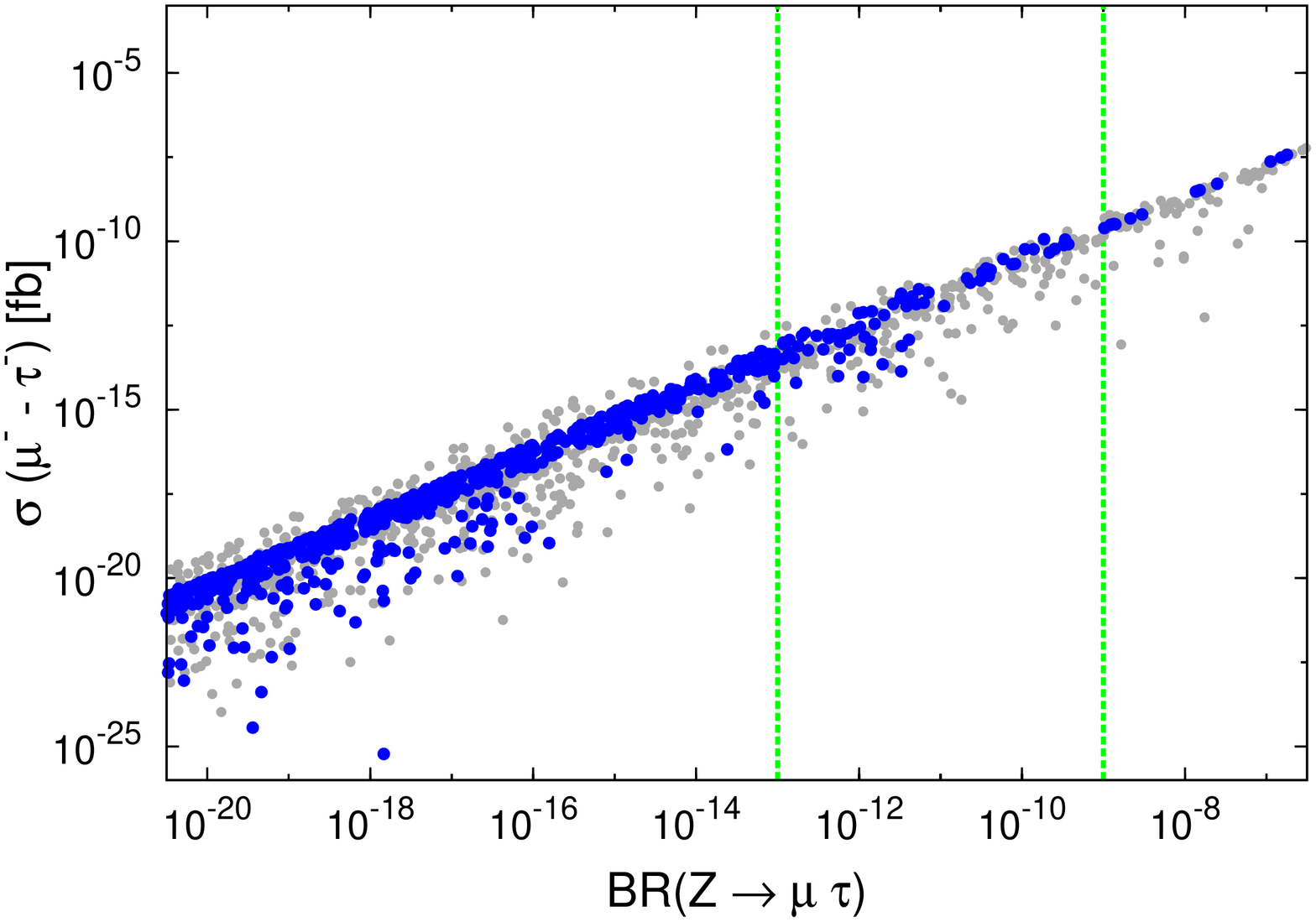,
  width=60mm , angle=-90 } \\
\epsfig{file=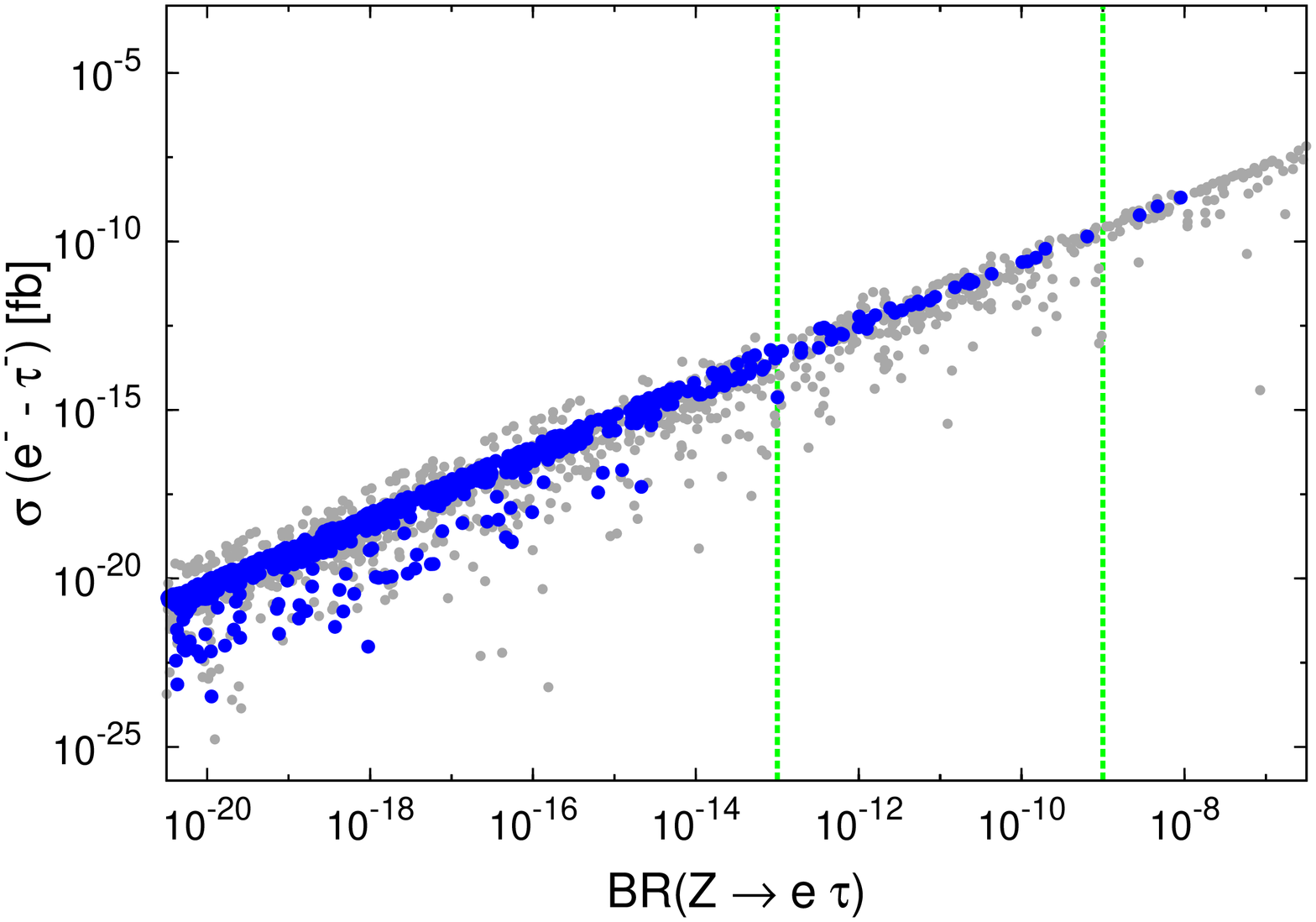,
  width=60mm , angle=-90 } 
&
\epsfig{file=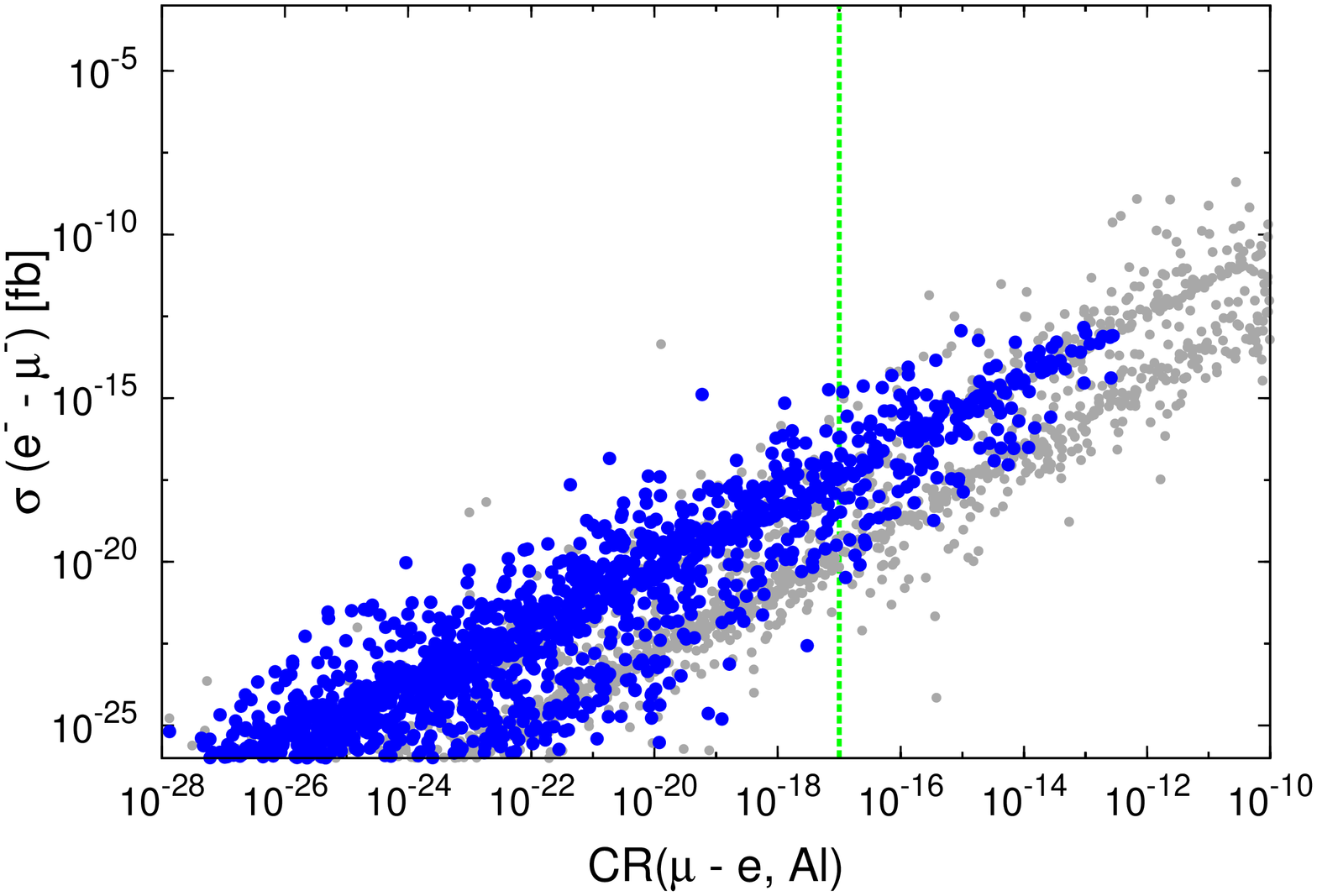,
  width=60mm, angle=-90 }
\end{tabular}
\end{center}
\caption{``3+1 model": correlation of cLFV in-flight 
cross sections with other cLFV observables. Upper panels: 
$\sigma (\mu \to \tau)$ vs. BR($\tau \to 3 \mu$) and 
BR($Z \to \mu \tau$). Lower panels: on the right 
$\sigma (e \to \tau)$ vs. BR($Z \to e \tau$); on the
left $\sigma (e \to \mu)$ vs. CR($\mu -e$, Al).
Blue coloured points comply with the different constraints 
discussed in Section~\ref{sec:models:constraints}; vertical full
(dashed) green lines denote, in each case, the corresponding 
current bounds (future sensitivities).}
\label{figs:sigma:eff.BR3ell.Zell} 
\end{figure}

\bigskip
As extensively discussed in the literature, the interplay of distinct
cLFV observables (arising from different sectors, and studied at
different energies and experimental setups) is a potentially 
powerful probe to test flavour violating extensions of the SM. 
For the case of our minimal framework - extending the SM with one
sterile fermion - we illustrate in Fig.~\ref{figs:sigma:eff.BR3ell.Zell}
the potential 
synergies between the in-flight conversion rate and other cLFV
observables, for which $Z$-penguin exchanges are known to
provide important (if not dominant) contributions: 
BR($\ell_j \to 3\ell_i$), BR($Z \to \ell_i \ell_j$), and - in the case
of $e-\mu$ conversion, CR($\mu -e$, N). 
As could be expected, there is a clear correlation between the
in-flight and both high-intensity and high-energy observables. 
Should one dispose of an unlimited number of leptons in the beam, 
the in-flight cLFV conversion could simultaneously probe - or even be
complementary to other low - and high-energy cLFV observables. 
Nevertheless, the small expected number of converted
leptons, for what are already optimistic beam configurations, dismisses the
latter possibilities.

To finalise the discussion, we briefly comment on the prospects for
this cLFV observable in well-motivated mechanisms of $\nu$ mass
generation, such as the ISS. 

\medskip
\noindent
{\it (3,3) Inverse seesaw realisation}

The numerical results for the ISS here displayed were obtained relying
on a random scan over the $9\times 9$ 
neutrino mass matrix (for a detailed
discussion of the numerical studies, see for example~\cite{Abada:2014cca});
we take the following ranges for the $M_R$ and $\mu_X$ matrices:
$0.5 \text{ GeV} \lesssim |(M_R)_{i}|  \lesssim 10^6 \text{ GeV}$ and
$0.01 \text{ eV} \lesssim
|(\mu_X)_{ij}|  \lesssim 1 \text{ MeV}$, with complex entries for the 
lepton number violating matrix $\mu_X$. In order to accommodate 
neutrino oscillation data, we use a modified  
Casas-Ibarra parametrisation~\cite{Casas:2001sr} 
for $Y^\nu$, with complex angles for the $R$ matrix which encodes 
the additional degrees of freedom (these are randomly 
varied in the interval $[0, 2\pi]$), always verifying that 
the Yukawa couplings are perturbative,
  i.e. $Y^\nu < 4 \pi$. All bounds referred to in
  Section~\ref{sec:models:constraints} are taken into account. 
For the purpose of this section, we consider a NH for the 
light neutrino spectrum. 

\begin{figure}
\begin{center}
\begin{tabular}{cc}
\epsfig{file=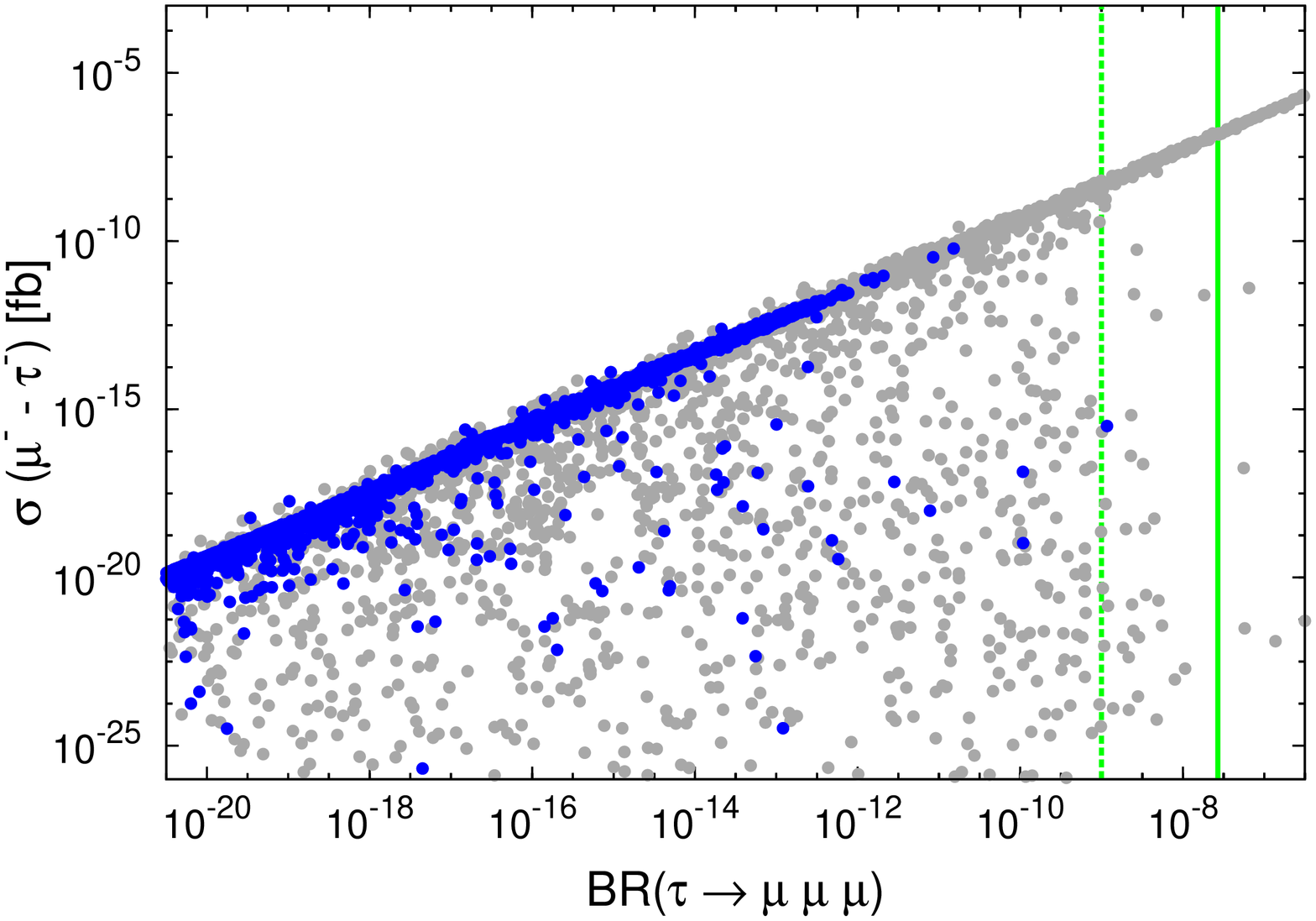,
  width=60mm, angle=-90 }
&
\epsfig{file=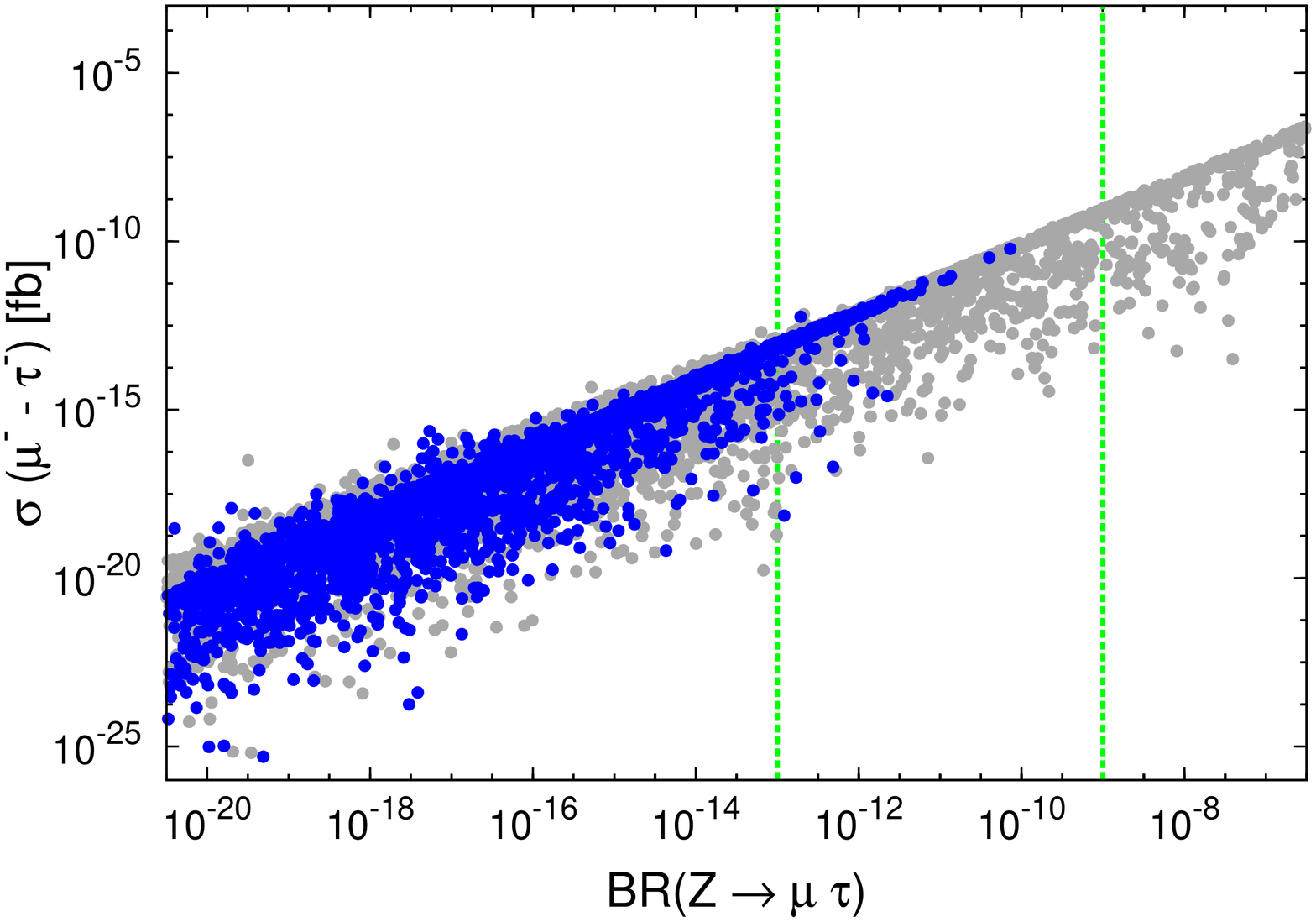,
  width=60mm , angle=-90 } \\
\epsfig{file=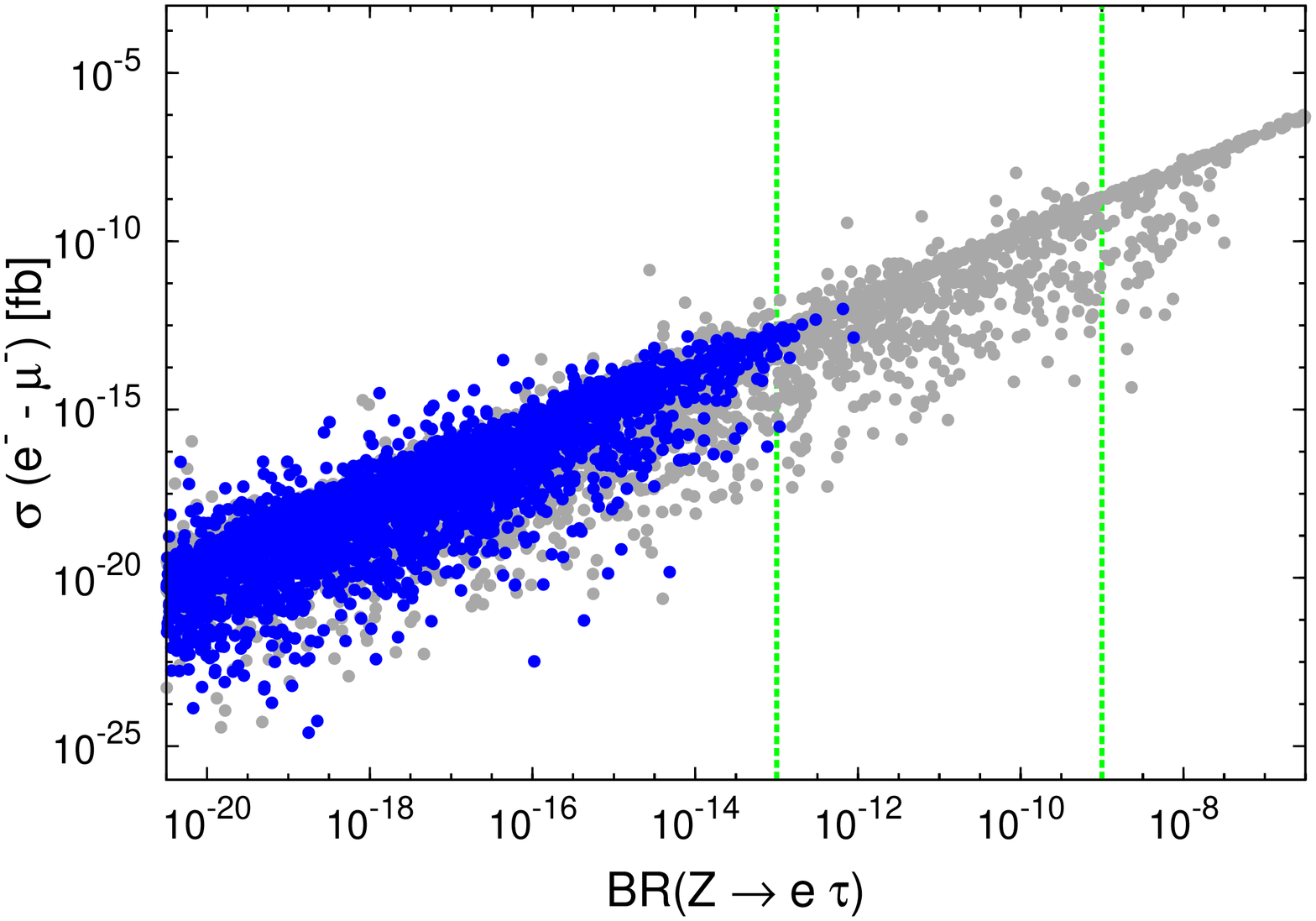,
  width=60mm , angle=-90 }  
&
\epsfig{file=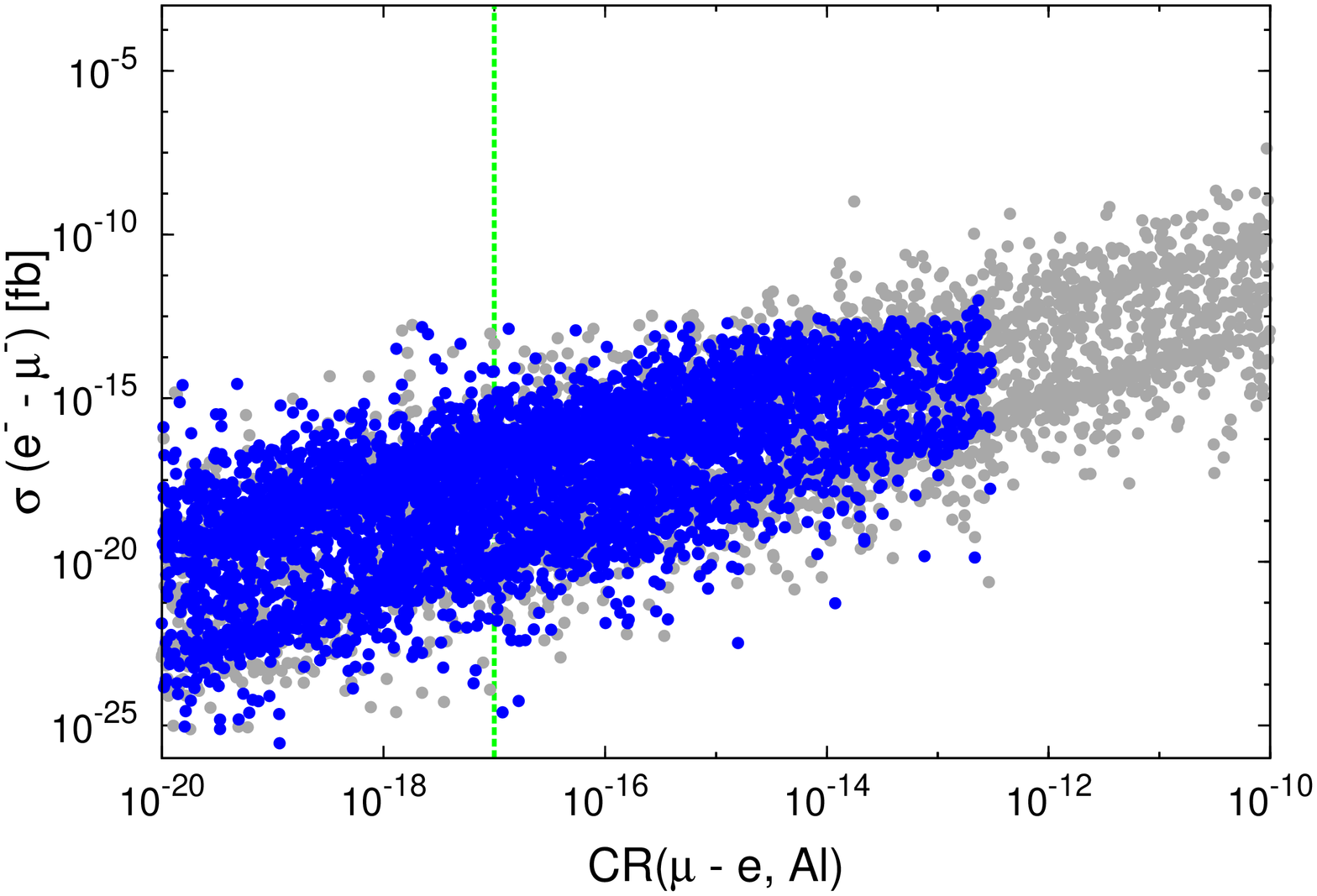,
  width=60mm, angle=-90 }
\end{tabular}
\end{center}
\caption{ISS realisation: correlation of cLFV in-flight
cross section with other cLFV observables. Upper panels: 
$\sigma (\mu n \to \tau n)$ vs. BR($\tau \to 3 \mu$) and 
BR($Z \to \mu \tau$). Lower panels: on the right 
$\sigma (e n \to \tau n)$ vs. BR($Z \to e \tau$); on the
left $\sigma (e n \to \mu n)$ vs. CR($\mu -e$, Al).
Line and colour code as in Fig.~\ref{figs:sigma:eff.BR3ell.Zell}.}
\label{figs:sigma.Q2:ISS.BR3ell.Zell} 
\end{figure}

We illustrate the synergy between the in-flight
conversion and other cLFV observables in the framework of the (3,3)
ISS; the distinct panels of Fig.~\ref{figs:sigma.Q2:ISS.BR3ell.Zell}
summarise a study similar to that displayed in
Fig.~\ref{figs:sigma:eff.BR3ell.Zell}.  
 
The summary of the ISS prospects, collected in
Fig.~\ref{figs:sigma.Q2:ISS.BR3ell.Zell} confirms what had been
previously found in studies of other cLFV observables 
(among them 3-body decays, conversion in Nuclei, or cLFV 
$Z$ decays): although such an ISS realisation can in principle account
for sizeable values of the in-flight cLFV conversion, experimental
bounds preclude the associated regimes. Fot instance, the maximal
expected values for the $\mu \to \tau$ integrated cross section does
not exceed $\sigma (\mu n \to \tau n) \lesssim \mathcal{O}(10^{-10})$
(while in the simple ``3+1 toy model'' one could have regions with 
$\sigma (\mu n \to \tau n)$ above 
$\mathcal{O}(10^{-8})$. 

While the simple ``3+1 toy
model'' (in which the active-sterile mixing is only constrained from 
experimental bounds), in the 
(3,3) ISS realisation the flavour violating structures 
(i.e., the Yukawa couplings and the LNV $\mu_X$ matrix) 
reflect correlations which are a consequence of necessarily  
accommodating oscillation data.  
Unlike the simple ``3+1 toy model'', which allowed to independently
explore different directions in flavour space (and thus, for example,  
evade $\mu-e$ sector constraints while enhancing $\mu-\tau$ flavour
violation), the ISS thus offers a far more constrained scenario.
We do not dismis that special textures - i.e., strongly suppressing
  mixings subject to the most stringent experimental bounds, while
  enhancing those which play a leading r\^ole in the observable -
  could account for higher
  values~\cite{Arganda:2014dta,Arganda:2015ija}. Nevertheless, these
  are somewhat fine-tuned constructions, 
  which we will not pursue in the present analysis.

\bigskip
\noindent
{\it $\nu$MSM}

We have also numerically explored the prospects of the $\nu$MSM;
despite the additional degrees of freedom 
 - in particular three new mixing angles $\theta_{45,46,56}$  
(other than the new Dirac and Majorana CP violating phases) -  
the allowed  
$\nu$MSM parameter space~\cite{Canetti:2012kh} leads to very poor
results for cLFV observables, with maximal values of the cLFV
in-flight cross sections many orders of magnitude below those arising
in the framework of the (3,3) ISS realisation above discussed. 
We notice that due to the very low scale of the new states (typically
below 100~GeV), which are accompanied by not excessively large
mixings, 
the general prospects of the $\nu$MSM for cLFV are not as appealing as
those of other low-scale seesaw realisations (see, 
for example,~\cite{Abada:2014cca, Abada:2015oba, Canetti:2013qna}), a
direct consequence of the size of its intrinsic sources of flavour
violation.

\section{Concluding remarks}
In the past years, 
charged lepton flavour violating observables have gained an increasing
interest stemming from their potential to probe scenarios of New
Physics, even those whose typical scales lie beyond collider reach. 
In view of upcoming facilities, which are expected to operate with intense
lepton beams (for example those dedicated to high-intensity cLFV
searches as COMET, NA64, future neutrino factories, or even a Muon
Collider), in-flight lepton flavour conversions occurring when the
intense beams hit a fixed target, are  potentially interesting 
cLFV observables.

In this study we have thus revisited cLFV in-flight conversion, 
$e + N \to \mu +N$, $e + N \to \tau +N$ and $\mu + N \to \tau +N$, 
focusing on elastic interactions with a nucleus $N$ (considering
moderately energetic beams, with an energy not far from the
kinematical threshold). We have studied the different contributions to
the differential cross sections, and our findings concerning the  
derivation of the leptonic and hadronic tensors are in agreement
with those of Ref.~\cite{Liao:2015hqu}. Motivated by classes of NP models
in which cLFV processes occur at higher order, we have moreover
focused on the dipole and $Z$-penguin contributions to the in-flight 
cLFV conversion. 

After a general discussion of the observable, we carried a 
thorough phenomenological analysis 
in the framework of minimal SM extensions via sterile
neutrinos, in which $Z$-penguin transitions do indeed dominate over
the dipole contributions (and box diagrams as well).  
Although such minimal frameworks do offer the possibility
to have sizeable values for the cross sections, $\sigma (\ell_i \to
\ell_j)$, these values are precluded due to the stringent bounds
arising from a number of other cLFV observables. 
Particularly constraining are
those observables in which the $Z$-penguin contributions also play 
a relevant r\^ole - among them BR($\mu \to 3 e$), BR($\tau \to 3 e$), 
BR($\tau \to 3 \mu$), and CR($\mu-e$, Au). 
Once the latter bounds are taken into account, the distinct cross
sections are strongly reduced - at most one expects values of
$\mathcal{O}(10^{-8}\text{ fb})$, for the case of $\mu -\tau$
conversion (for which the associated low-energy cLFV constraints are
less stringent). 
Even when assuming the possibility of very intense
lepton beams, our study suggests that the 
expected number of conversions lies beyond
experimental sensitivity (below $\mathcal{O}(10^{-2}\text{
  events/year})$).
 
Other theoretical frameworks relying on extensions of
the SM via several sterile fermions were found to lead to similar
(or even worse) prospects: studies of the in-flight cLFV observables in
complete
models as the (3,3) ISS realisation or the 
$\nu$MSM, were carried, and our findings confirmed that such
frameworks indeed accounted for smaller predictions to the distinct
observables than what is found in the framework of the simple 
``3+1 model''.

Albeit the results here obtained concern minimal SM extensions via sterile
neutrinos, the strong correlation between the in-flight conversion and
the cLFV observables which preclude its observability should be common
to other NP constructions exhibiting similar features. This is the
case of minimal (constrained) SUSY models, where there is typically a strong
correlation between radiative decays and the $\gamma$-penguins
providing the dominant contributions to 3-body decays; in this sense,
our findings confirm those of~\cite{Blazek:2004cg}
which pointed out that former bounds on $\mu \to e \gamma$ already forbade
SUSY contributions to  $\sigma (e \to \mu)$ larger than $10^{-8}$~fb.

It is also worth considering the possibility of having additional
sterile states: if on the one hand this might contribute to enhance
the $\ell_i \to \ell_j$ cross sections (via a multiplicative factor,
thus leading at most to a single order of magnitude enhancement), the
additional states would also contribute to the other cLFV observables,
so that one does not expect an overall improvement. Likewise, a study
in the DIS limit should not qualitatively change the general results
here derived.  

Should experimental searches for the in-flight cLFV conversion 
observable be carried in
the future, and should an event be observed, then another source of flavour
violation, different from - or in addition to - those present in
minimal SM extensions via sterile fermions must be necessarily
present. Moreover, available (phenomenological) 
results would suggest that such a NP
model would likely exhibit a smaller degree of correlation between
different cLFV observables (as is the case of leptoquark models): for example, 
some transitions occurring at tree-level,
while others being mediated via higher order exchanges. 

Finally, and as in the case of any cLFV observable, the experimental
observation of  
the in-flight cLFV conversion (as could happen in the near future at
NA64~\cite{SPS:NA64, Gninenko:private}), 
would clearly signal the presence of New
Physics, and allow selecting classes of models (other than those
here discussed) which could account for it.

\section*{Acknowledgements}
We are grateful to S. Gninenko for interesting remarks and contributions.
This work was done within the framework of the European Union's Horizon 2020
research and innovation programme under
the Marie Sklodowska-Curie grant agreements No 690575 and No 674896. 
VDR acknowledges support by the Spanish MINECO through the project
FPA2012-31880 (P.I. E. Alvarez Vazquez).

\appendix

\section{Nuclear and leptonic tensors}\label{appendix:dsigma:computation}
We describe the most relevant elements leading to the computation of
both the photon- and $Z$-mediated interactions. 

\subsection{Photonic interaction}
The hadronic tensor relevant for the photon mediated on-target
conversion, as given in Eq.~(\ref{eq:photon:dsigma.dQ2:photon}), 
can be cast as
\begin{align}\label{eq:photon:dsigma.dQ2:photon:Hmunu}
H^\gamma_{\mu \nu}\, =& -(\eta_{\mu \nu} -q_\mu q_\nu/q^2)\, W_1 
\, +\, \frac{1}{M^2_T}(p_\mu -q_\mu\, p.q/q^2)\,
(p_\nu -q_\nu\, p.q/q^2)\,W_2\,,
\end{align}
where, and for a spin $1/2$ target, one has 
\begin{equation}\label{eq:photon:dsigma.dQ2:photon:W1W2}
W_1\, =\, \frac{Q^2}{4\, M^2_T} (F_1+F_2)^2\,, \quad \quad
W_2\, =\, F_1^2 \,+ \, \frac{Q^2}{4\, M^2_T} F_2^2\,,
\end{equation}
with $F_{1,2}$ the Dirac and Pauli form factors, which in our analysis
refer to the nucleon form factors, $F_i^{p,n}$. In agreement 
with~\cite{Liao:2015hqu}, one can write the latter as:
\begin{align} 
& F_1^{p (n)} (Q^2)\, =\, \frac{1}{1+Q^2/4 M_N^2 }\,
\left[
\frac{1\, (0)}{1+Q^2/4 M_V^2} + \frac{Q^2}{4 M_N^2} 
\frac{\mu_p \, (\mu_n)}{1+Q^2/4 M_V^2}\right] \,, \nonumber \\
& F_2^{p (n)} (Q^2)\, =\, \frac{1}{1+Q^2/4 M_N^2 }\,
\left[\frac{\mu_p - 1 \, (\mu_n)}{1+Q^2/4 M_V^2}\right] \,,
\end{align}
where $M_N$ denotes the nucleon mass ($M_{p,n}$), and $M_V$ the
relevant scale for the interaction, $M_V~=~M_W$, and $\mu_{p,n}$ the
total magnetic moments, respectively $\mu_{p (n)} = 2.79~(-1.91)~e /2
M_{p(n)}$. 

\medskip
Likewise, the leptonic tensor also present in
Eq.~(\ref{eq:photon:dsigma.dQ2:photon})
can be expressed in terms of momenta 
as\footnote{We adopt a similar notation to that 
of Ref.~\cite{Liao:2015hqu}, the results of which we agree with.}
\begin{align}\label{eq:photon:dsigma.dQ2:photon:Lmunu:app}
L^\gamma_{\mu \nu}\, =& -2 \left [m_\ell^2\,(m_\ell^2 -q^2)\, 
(\eta_{\mu \nu} -q_\mu q_\nu/q^2) \, +\,  
4\, q^2\, (k^\mu - q^\mu \,k.q/q^2)\,
(k^\nu - q^\nu \,k.q/q^2)\right]\,,
\end{align}
in which $m_\ell$ denotes the mass of the final state (heavier)
lepton. 

Bringing all the elements together, the final expression for the
photon contribution to the differential cross section is given by 
\begin{align}
\left. \frac{d \sigma^{i\to j}}{d Q^2} \right|_{\gamma}\, =\, &
\frac{2 \pi \,Z^2 \,\alpha^2}{E^2 \,Q^4} \, L^\gamma_{ij}\, \left \{
W_1\, (Q^2 + m^2_{\ell_j})\, (2 m^2_{\ell_j} - Q^2) \,+\,
\right.
\nonumber \\
& + \, \left.
\frac{W_2}{M^2_T}\left (
4\, Q^2\, (p. k)^2 \, +\, 
(Q^2 + m^2_{\ell_j})\, \left[(p . q)^2 \, -\, 
4\, p. q \, p. k\, +\, M_T^2 \,m^2_{\ell_j}\right] \right)
\right \}\,,
\end{align}
with $L^\gamma_{ij}$ given in Eq.~(\ref{eq:photon:dsigma.dQ2:photon:Lij}).

\subsection{$Z$-mediated interaction}
Assuming the case of unpolarised lepton beams, 
the leptonic tensor entering in the $Z$-interaction contribution to
the differential cross section (see Eq.~(\ref{eq:photon:dsigma.dQ2:Z}))
can be written as 
\begin{align}\label{eq:photon:dsigma.dQ2:Z:Lmunu:app}
L^Z_{\mu \nu}\, =& \,16 \, \left(
k^\mu k^{\prime \nu} + k^\nu k^{\prime \mu} - k.k^\prime \eta^{\mu \nu}
+ i \, \varepsilon^{\mu \nu \rho \sigma} k_\rho q_\sigma \right)\,.
\end{align}
The hadronic tensor can be in general cast in terms of six
dimensionless structure functions as 
\begin{align}\label{eq:photon:dsigma.dQ2:Z:Hmunu}
H^Z_{\mu \nu}\, =& -\eta_{\mu \nu}\, W_1 +\frac{p_\mu p_\nu}{M^2_T}\,W_2
\pm i \, \varepsilon_{\mu \nu \rho \sigma} \frac{p^\rho q^\sigma}{2 M^2_T}\,W_3
+ \frac{q_\mu q_\nu}{M^2_T}\,W_4 + 
\frac{p_\mu q_\nu + p_\nu q_\mu}{2 M^2_T}\,W_5 +
i \,  \frac{p_\mu q_\nu - p_\nu q_\mu}{2 M^2_T}\,W_6\,,
\end{align}
where in the above equation the $\pm$ corresponds 
to having a lepton (or antilepton)
conversion. The different expressions for the
structure functions $W_i (Q^2)$ can be found in
Ref.~\cite{Liao:2015hqu}, with which we agree after an independent
derivation.  

As above, the contraction of both leptonic and hadronic tensors leads to
the following differential cross section for 
$Z$-mediated contribution, which we have used throughout the analysis, 
\begin{align}
\left. \frac{d \sigma^{i\to j}}{d Q^2} \right|_{Z}\, =\, &
\frac{G_F^2}{2 \pi \, E^2}\, L^Z_{ij}\, \left \{ \left[
(Q^2 + m^2_{\ell_j})\,(W_1 - \frac{1}{2} W_2)\, +\, 
\frac{p.k}{M^2_N}\, (2 \, p.k-Q^2)\, W_2 \, +\, 
\right. \right.
\nonumber \\
& + \, \left.\left.
\frac{1}{2}(Q^2 + m^2_{\ell_j})\, \frac{m^2_{\ell_j}}{M^2_N}\, W_4 \,
-\, 
(p.k)\,\frac{m^2_{\ell_j}}{M^2_N}\, W_5 \right]\, 
\pm\,
\frac{Q^2}{4 M^2_N}\, (4 \,p.k - Q^2 - m^2_{\ell_j})\, W_3
\right \}\,,
\end{align}
with $L^Z_{ij}$ has been given in Eq.~(\ref{eq:photon:dsigma.dQ2:Z:Lijpm}).

\section{cLFV form factors}\label{appendix:cLFV:formulae}
The relevant form factors for the computation of the diagrams of
Fig.~\ref{fig:diagrams} are given
by~\cite{Alonso:2012ji,Ilakovac:1994kj,Ma:1979px,Gronau:1984ct}:
\begin{eqnarray}
\label{eq:FF}
G^{\ell m }_\gamma &=& \sum_{j=1}^{3 + n_S} {\bf U}_{mj}{\bf U}^*_{\ell
  j} G_\gamma(x_j)\,,  \quad \quad 
F^{\ell m }_\gamma = 
\sum_{j=1}^{3 + n_S} {\bf U}_{mj}{\bf U}^*_{\ell
  j} F_\gamma(x_j)\,, \nonumber \label{Fmue}\\ 
F^{\ell m }_Z &=& \sum_{j,k=1}^{3 + n_S} {\bf U}_{mj}{\bf U}^*_{\ell k}
\left(\delta_{jk} F_Z(x_j) + {\bf C}_{jk} G_Z(x_j,x_k) + {\bf
  C}^*_{jk} H_Z(x_j,x_k)   \right)\,,   
\end{eqnarray}
where $x_i = \frac{m^2_{\nu_i}}{m_W^2}$ carries the neutrino mass 
dependency and ${\bf C}$ has been defined in
Eq.~(\ref{eq:Cmatrix:def}).

The loop functions entering the previous form factors are defined
as~\cite{Alonso:2012ji,Ilakovac:1994kj,Ma:1979px,Gronau:1984ct}: 
\begin{align}
& F_Z(x)\,=\, -\frac{5x}{2\,(1-x)}-\frac{5x^2}{2\,(1-x)^2}\ln x \, , \nonumber
\\  
& G_Z(x,y)\,=\, -\frac{1}{2(x-y)}\left[	\frac{x^2\,(1-y)}{1-x}\ln x -
  \frac{y^2\,(1-x)}{1-y}\ln y	\right]\, ,  \nonumber \\ 
& H_Z(x,y)\,=\,  \frac{\sqrt{x\,y}}{4\,(x-y)}\left[	
\frac{x^2-4x}{1-x}\ln
  x - \frac{y^2-4y}{1-y}\ln y	\right] \, , \nonumber \\ 
& F_\gamma(x)\,= \,	\frac{x\,(7x^2-x-12)}{12\,(1-x)^3} -
\frac{x^2\,(x^2-10x+12)}{6\,(1-x)^4} \ln x	\, , \nonumber \\ 
& G_\gamma(x)\,=  \,  -\frac{x\,(2x^2+5x-1)}{4\,(1-x)^3} -
\frac{3x^3}{2\,(1-x)^4} \ln x \,.\label{Ggamma}			
 \end{align}

The contributions to different cLFV observables such as radiative
and 3-body decays, conversion in Nuclei, or FV $Z$ decays,  
which have been
evaluated and analysed in the present work (including the
relevant loop
functions~\cite{Alonso:2012ji,Ilakovac:1994kj,Ma:1979px,Gronau:1984ct}),
have been discussed in previous studies 
(see, for
example,~\cite{Alonso:2012ji,Ilakovac:1994kj,Abada:2014cca,Abada:2015oba}), 
and we will not include them here.


{\small
\begin{thebibliography}{99}


  \bibitem{TheMEG:2016wtm}
  A.~M.~Baldini {\it et al.} [MEG Collaboration],
  Eur.\ Phys.\ J.\ C {\bf 76} (2016) no.8,  434
  [arXiv:1605.05081 [hep-ex]].
  
  \bibitem{Baldini:2013ke}
  A.~M.~Baldini, F.~Cei, C.~Cerri, S.~Dussoni, L.~Galli, M.~Grassi,
  D.~Nicolo and F.~Raffaelli {\it et al.}, 
  arXiv:1301.7225 [physics.ins-det].
  
 \bibitem{Aubert:2009ag}
  B.~Aubert {\it et al.}  [BaBar Collaboration],
  Phys.\ Rev.\ Lett.\  {\bf 104} (2010) 021802
  
  \bibitem{Aushev:2010bq}
  T.~Aushev, W.~Bartel, A.~Bondar, J.~Brodzicka, T.~E.~Browder,
  P.~Chang, Y.~Chao and K.~F.~Chen {\it et al.}, 
  ``Physics at Super B Factory,''
  arXiv:1002.5012 [hep-ex].
  

  \bibitem{Bellgardt:1987du}
  U.~Bellgardt {\it et al.}  [SINDRUM Collaboration],
  Nucl.\ Phys.\ B {\bf 299} (1988) 1.
  
  \bibitem{Blondel:2013ia}
 A.~Blondel, A.~Bravar, M.~Pohl, S.~Bachmann, N.~Berger, M.~Kiehn,
 A.~Schoning and D.~Wiedner {\it et al.}, 
  ``Research Proposal for an Experiment to Search for the Decay $\mu \to eee$,''
  arXiv:1301.6113 [physics.ins-det].
   
 \bibitem{Hayasaka:2010np}
  K.~Hayasaka, K.~Inami, Y.~Miyazaki, K.~Arinstein, V.~Aulchenko,
  T.~Aushev, A.~M.~Bakich and A.~Bay {\it et al.}, 
  Phys.\ Lett.\ B {\bf 687} (2010) 139
  [arXiv:1001.3221 [hep-ex]].

\bibitem{Bertl:2006up}
  W.~H.~Bertl {\it et al.}  [SINDRUM II Collaboration],
  Eur.\ Phys.\ J.\ C {\bf 47} (2006) 337.
    
\bibitem{Aoki:2012zza}
  M.~Aoki [DeeMe Collaboration],
  AIP Conf.\ Proc.\  {\bf 1441} (2012) 599.

\bibitem{Carey:2008zz}
  R.~M.~Carey {\it et al.} [Mu2e Collaboration],
  FERMILAB-PROPOSAL-0973.
  
\bibitem{Cui:2009zz}
  Y.~G.~Cui {\it et al.} [COMET Collaboration],
  KEK-2009-10.
  
\bibitem{Kuno:2013mha}
  Y.~Kuno [COMET Collaboration],
  PTEP {\bf 2013} (2013) 022C01. 

\bibitem{Koike:2010xr}
  M.~Koike, Y.~Kuno, J.~Sato and M.~Yamanaka,
  Phys.\ Rev.\ Lett.\  {\bf 105} (2010) 121601
  [arXiv:1003.1578 [hep-ph]].
  
\bibitem{Uesaka:2016vfy}
  Y.~Uesaka, Y.~Kuno, J.~Sato, T.~Sato and M.~Yamanaka,
  Phys.\ Rev.\ D {\bf 93} (2016) no.7,  076006
  [arXiv:1603.01522 [hep-ph]].

\bibitem{Littenberg:2000fg}
  L.~S.~Littenberg and R.~Shrock,
  Phys.\ Lett.\ B {\bf 491} (2000) 285
  [hep-ph/0005285].

\bibitem{Geib:2016atx}
  T.~Geib, A.~Merle and K.~Zuber,
  Phys.\ Lett.\ B {\bf 764} (2017) 157
  [arXiv:1609.09088 [hep-ph]].

\bibitem{Berryman:2016slh}
  J.~M.~Berryman, A.~de Gouvea, K.~J.~Kelly and A.~Kobach,
  ``On Lepton-Number-Violating Searches for Muon to Positron Conversion,''
  arXiv:1611.00032 [hep-ph].

\bibitem{Geib:2016daa}
  T.~Geib and A.~Merle,
  arXiv:1612.00452 [hep-ph].

\bibitem{Gninenko:2001id}
  S.~N.~Gninenko, M.~M.~Kirsanov, N.~V.~Krasnikov and V.~A.~Matveev,
  Mod.\ Phys.\ Lett.\ A {\bf 17} (2002) 1407
  doi:10.1142/S0217732302007855
  [hep-ph/0106302].

\bibitem{Sher:2003vi}
  M.~Sher and I.~Turan,
  Phys.\ Rev.\ D {\bf 69} (2004) 017302
  [hep-ph/0309183].

\bibitem{Gonderinger:2010yn}
  M.~Gonderinger and M.~J.~Ramsey-Musolf,
  JHEP {\bf 1011} (2010) 045
   Erratum: [JHEP {\bf 1205} (2012) 047]
  [arXiv:1006.5063 [hep-ph]].

\bibitem{Kanemura:2004jt}
  S.~Kanemura, Y.~Kuno, M.~Kuze and T.~Ota,
  Phys.\ Lett.\ B {\bf 607} (2005) 165
  [hep-ph/0410044].

\bibitem{SPS:NA64}
  S.~N.~Gninenko,
  Phys.\ Rev.\ D {\bf 89} (2014) no.7,  075008
  [arXiv:1308.6521 [hep-ph]]; 
  S.~Andreas {\it et al.},
  ``Proposal for an Experiment to Search for Light Dark Matter at the SPS,''
  arXiv:1312.3309 [hep-ex]. See also https://na64.web.cern.ch/

\bibitem{Blazek:2004cg}
  T.~Blazek and S.~F.~King,
  ``Electron to muon conversion in electron-nucleus scattering as a
  probe of supersymmetry,'' 
  hep-ph/0408157.

\bibitem{Diener:2004kq}
  K.~P.~O.~Diener,
  Nucl.\ Phys.\ B {\bf 697} (2004) 387
  [hep-ph/0403251].

\bibitem{Bolanos:2012zd}
  A.~Bolanos, A.~Fernandez, A.~Moyotl and G.~Tavares-Velasco,
  Phys.\ Rev.\ D {\bf 87} (2013) no.1,  016004
  [arXiv:1212.0904 [hep-ph]].

\bibitem{Liao:2015hqu}
  W.~Liao and X.~H.~Wu,
  Phys.\ Rev.\ D {\bf 93} (2016) no.1,  016011
  [arXiv:1512.01951 [hep-ph]].

\bibitem{Bernstein:2013hba}
  R.~H.~Bernstein and P.~S.~Cooper,
  Phys.\ Rept.\  {\bf 532} (2013) 27
  [arXiv:1307.5787 [hep-ex]].

\bibitem{Ma:1979px}
 E.~Ma and A.~Pramudita,
 Phys.\ Rev.\ D {\bf 22} (1980) 214.

\bibitem{Gronau:1984ct}
  M.~Gronau, C.~N.~Leung and J.~L.~Rosner,
  Phys.\ Rev.\ D {\bf 29} (1984) 2539.

\bibitem{Ilakovac:1994kj}
  A.~Ilakovac and A.~Pilaftsis,
  Nucl.\ Phys.\ B {\bf 437} (1995) 491
  [hep-ph/9403398].

\bibitem{Deppisch:2004fa}
  F.~Deppisch and J.~W.~F.~Valle,
  Phys.\ Rev.\ D {\bf 72} (2005) 036001
  [hep-ph/0406040].

\bibitem{Deppisch:2005zm}
 F.~Deppisch, T.~S.~Kosmas and J.~W.~F.~Valle,
 Nucl.\ Phys.\ B {\bf 752} (2006) 80
 [hep-ph/0512360].

\bibitem{Dinh:2012bp}
 D.~N.~Dinh, A.~Ibarra, E.~Molinaro and S.~T.~Petcov,
Decays and TeV Scale See-Saw Scenarios of Neutrino Mass Generation,''
 JHEP {\bf 1208} (2012) 125
  [Erratum-ibid.\  {\bf 1309} (2013) 023]
 [arXiv:1205.4671 [hep-ph]].

\bibitem{Alonso:2012ji}
 R.~Alonso, M.~Dhen, M.~B.~Gavela and T.~Hambye,
 JHEP {\bf 1301} (2013) 118
 [arXiv:1209.2679 [hep-ph]].

\bibitem{Arganda:2014dta}
 E.~Arganda, M.~J.~Herrero, X.~Marcano and C.~Weiland,
 Phys.\ Rev.\ D {\bf 91} (2015) no.1,  015001
 [arXiv:1405.4300 [hep-ph]].

\bibitem{Abada:2014kba}
  A.~Abada, M.~E.~Krauss, W.~Porod, F.~Staub, A.~Vicente and C.~Weiland,
  JHEP {\bf 1411} (2014) 048
  [arXiv:1408.0138 [hep-ph]].

 \bibitem{Abada:2014cca}
  A.~Abada, V.~De Romeri, S.~Monteil, J.~Orloff and A.~M.~Teixeira,
  JHEP {\bf 1504} (2015) 051
  [arXiv:1412.6322 [hep-ph]].

\bibitem{Abada:2015zea}
  A.~Abada, D.~Becirevic, M.~Lucente and O.~Sumensari,
  Phys.\ Rev.\ D {\bf 91} (2015) no.11,  113013
  [arXiv:1503.04159 [hep-ph]].

\bibitem{Abada:2015oba}
  A.~Abada, V.~De Romeri and A.~M.~Teixeira,
  JHEP {\bf 1602} (2016) 083
  [arXiv:1510.06657 [hep-ph]].

\bibitem{DeRomeri:2016gum}
 V.~De Romeri, M.~J.~Herrero, X.~Marcano and F.~Scarcella,
 ``Lepton flavor violating Z decays: A promising window to low scale
 seesaw neutrinos,'' 
 arXiv:1607.05257 [hep-ph].

\bibitem{Akhmedov:1995vm}
  E.~K.~Akhmedov, M.~Lindner, E.~Schnapka and J.~W.~F.~Valle,
  Phys.\ Rev.\ D {\bf 53} (1996) 2752
  [hep-ph/9509255].

\bibitem{Malinsky:2005bi}
  M.~Malinsky, J.~C.~Romao and J.~W.~F.~Valle,
  Phys.\ Rev.\ Lett.\  {\bf 95} (2005) 161801
  [hep-ph/0506296].

\bibitem{Mohapatra:1986bd} 
  R.~N.~Mohapatra and J.~W.~F.~Valle,
  Phys.\ Rev.\ D {\bf 34} (1986) 1642.

\bibitem{Asaka:2005an}
  T.~Asaka, S.~Blanchet and M.~Shaposhnikov,
  Phys.\ Lett.\ B {\bf 631} (2005) 151
  [hep-ph/0503065].

\bibitem{Schechter:1980gr}
  J.~Schechter and J.~W.~F.~Valle,
  Phys.\ Rev.\ D {\bf 22} (1980) 2227.

\bibitem{Chanowitz:1978mv}
  M.~S.~Chanowitz, M.~A.~Furman and I.~Hinchliffe,
  Nucl.\ Phys.\ B {\bf 153} (1979) 402.

\bibitem{Durand:1989zs}
  L.~Durand, J.~M.~Johnson and J.~L.~Lopez,
  Phys.\ Rev.\ Lett.\  {\bf 64} (1990) 1215.

\bibitem{Korner:1992an}
  J.~G.~Korner, A.~Pilaftsis and K.~Schilcher,
  Phys.\ Lett.\ B {\bf 300} (1993) 381
  [hep-ph/9301290].

\bibitem{Bernabeu:1993up}
  J.~Bernabeu, J.~G.~Korner, A.~Pilaftsis and K.~Schilcher,
  Phys.\ Rev.\ Lett.\  {\bf 71} (1993) 2695
  [hep-ph/9307295].
    
\bibitem{Fajfer:1998px}
  S.~Fajfer and A.~Ilakovac,
  Phys.\ Rev.\ D {\bf 57} (1998) 4219.

\bibitem{Ilakovac:1999md}
  A.~Ilakovac,
  Phys.\ Rev.\ D {\bf 62} (2000) 036010
  [hep-ph/9910213].



 \bibitem{Tortola:2012te} 
D.~V.~Forero, M.~Tortola and J.~W.~F.~Valle,
Phys.\ Rev.\ D {\bf 86} (2012) 073012 
[arXiv:1205.4018 [hep-ph]].
  
\bibitem{Fogli:2012ua}
G.~L.~Fogli, E.~Lisi, A.~Marrone, D.~Montanino, A.~Palazzo and A.~M.~Rotunno,
Phys.\ Rev.\ D {\bf 86} (2012) 013012
[arXiv:1205.5254 [hep-ph]].

\bibitem{GonzalezGarcia:2012sz}
M.~C.~Gonzalez-Garcia, M.~Maltoni, J.~Salvado and T.~Schwetz,
JHEP {\bf 1212} (2012) 123
[arXiv:1209.3023 [hep-ph]].

\bibitem{Forero:2014bxa}
  D.~V.~Forero, M.~Tortola and J.~W.~F.~Valle,
  Phys.\ Rev.\ D {\bf 90} (2014) 093006
  [arXiv:1405.7540 [hep-ph]].

\bibitem{nufit}
See also http://www.nu-fit.org/

\bibitem{Gonzalez-Garcia:2014bfa}
  M.~C.~Gonzalez-Garcia, M.~Maltoni and T.~Schwetz,
  JHEP {\bf 1411} (2014) 052
  [arXiv:1409.5439 [hep-ph]].
 
\bibitem{Esteban:2016qun}
  I.~Esteban, M.~C.~Gonzalez-Garcia, M.~Maltoni, I.~Martinez-Soler and
  T.~Schwetz, 
  ``Updated fit to three neutrino mixing: exploring the
  accelerator-reactor complementarity,'' 
  arXiv:1611.01514 [hep-ph].
  
\bibitem{Antusch:2008tz} 
S.~Antusch, J.~P.~Baumann and E.~Fernandez-Martinez,
Nucl.\ Phys.\ B {\bf 810} (2009) 369 
[arXiv:0807.1003 [hep-ph]].

\bibitem{Antusch:2014woa}
  S.~Antusch and O.~Fischer,
  JHEP {\bf 1410} (2014) 94
  [arXiv:1407.6607 [hep-ph]].

\bibitem{Blennow:2016jkn}
  M.~Blennow, P.~Coloma, E.~Fernandez-Martinez, J.~Hernandez-Garcia
  and J.~Lopez-Pavon, 
  ``Non-Unitarity, sterile neutrinos, and Non-Standard neutrino Interactions,''
  arXiv:1609.08637 [hep-ph].
  
  \bibitem{Akhmedov:2013hec}
  E.~Akhmedov, A.~Kartavtsev, M.~Lindner, L.~Michaels and J.~Smirnov,
  JHEP {\bf 1305} (2013) 081
  [arXiv:1302.1872 [hep-ph]].
  
\bibitem{Fernandez-Martinez:2015hxa}
  E.~Fernandez-Martinez, J.~Hernandez-Garcia, J.~Lopez-Pavon and M.~Lucente,
   JHEP {\bf 1510} (2015) 130 [arXiv:1508.03051 [hep-ph]].

\bibitem{Basso:2013jka}
L.~Basso, O.~Fischer and J.~J.~van der Bij,
Europhys.\ Lett.\  {\bf 105} (2014) 11001
[arXiv:1310.2057 [hep-ph]].

\bibitem{Abada:2013aba}
  A.~Abada, A.~M.~Teixeira, A.~Vicente and C.~Weiland,
  JHEP {\bf 1402} (2014) 091
  [arXiv:1311.2830 [hep-ph]].
  
\bibitem{Olive:2016xmw}
  C.~Patrignani {\it et al.} [Particle Data Group Collaboration],
  Chin.\ Phys.\ C {\bf 40} (2016) no.10, 100001.
  
\bibitem{BhupalDev:2012zg}
P.~S.~Bhupal Dev, R.~Franceschini and R.~N.~Mohapatra,
Phys.\ Rev.\ D {\bf 86} (2012) 093010
[arXiv:1207.2756 [hep-ph]].
 
\bibitem{Cely:2012bz}
C.~G.~Cely, A.~Ibarra, E.~Molinaro and S.~T.~Petcov,
Phys.\ Lett.\ B {\bf 718} (2013) 957
[arXiv:1208.3654 [hep-ph]].
  
\bibitem{Bandyopadhyay:2012px}
  P.~Bandyopadhyay, E.~J.~Chun, H.~Okada and J.~-C.~Park,
  JHEP {\bf 1301} (2013) 079
  [arXiv:1209.4803 [hep-ph]].
  
 \bibitem{Kusenko:2009up}
  A.~Kusenko,
  Phys.\ Rept.\  {\bf 481} (2009) 1
  [arXiv:0906.2968 [hep-ph]].

\bibitem{Atre:2009rg} 
  A.~Atre, T.~Han, S.~Pascoli and B.~Zhang,
  JHEP {\bf 0905}  (2009) 030
  [arXiv:0901.3589 [hep-ph]].

\bibitem{Goudzovski:2011tc} 
  E.~Goudzovski [NA48/2 and NA62 Collaborations],
  PoS EPS {\bf -HEP2011} (2011) 181 
  [arXiv:1111.2818 [hep-ex]].
  
\bibitem{Lazzeroni:2012cx} 
  C.~Lazzeroni {\it et al.}  [NA62 Collaboration],
  Phys.\ Lett.\ B {\bf 719} (2013) 326 
  [arXiv:1212.4012 [hep-ex]].
  
\bibitem{Naik:2009tk} 
  P.~Naik {\it et al.}  [CLEO Collaboration],
  Phys.\ Rev.\ D {\bf 80} (2009) 112004 
  [arXiv:0910.3602 [hep-ex]].
  
\bibitem{Li:2011nij} 
  H.~-B.~Li,
  ``Proceedings, 4th International Workshop on Charm Physics (Charm
  2010) : Beijing, China, October 21-24, 2010,'' 
  Int.\ J.\ Mod.\ Phys.\ Conf.\ Ser.\  {\bf 02} (2011).
  
\bibitem{Aubert:2007xj} 
  B.~Aubert {\it et al.}  [BaBar Collaboration],
  Phys.\ Rev.\ D {\bf 77} (2008) 011107.
  
\bibitem{Adachi:2012mm} 
  I.~Adachi {\it et al.}  [Belle Collaboration],
  Phys.\ Rev.\ Lett.\  {\bf 110}  (2013) 131801
  [arXiv:1208.4678 [hep-ex]].
  
  \bibitem{Abada:2012mc}
  A.~Abada, D.~Das, A.~M.~Teixeira, A.~Vicente and C.~Weiland,
  JHEP {\bf 1302} (2013) 048
  [arXiv:1211.3052 [hep-ph]].

\bibitem{Albert:2014awa}
  J.~B.~Albert {\it et al.} [EXO-200 Collaboration],
  Nature {\bf 510} (2014) 229
  [arXiv:1402.6956 [nucl-ex]].

 \bibitem{Benes:2005hn}
  P.~Benes, A.~Faessler, F.~Simkovic and S.~Kovalenko,
  Phys.\ Rev.\ D {\bf 71} (2005) 077901
  [hep-ph/0501295].  
  
  \bibitem{Blennow:2010th} 
  M.~Blennow, E.~Fernandez-Martinez, J.~Lopez-Pavon and J.~Menendez,
  JHEP {\bf 1007} (2010) 096 
  [arXiv:1005.3240 [hep-ph]].

  \bibitem{Abada:2014nwa}
  A.~Abada, V.~De Romeri and A.~M.~Teixeira,
  JHEP {\bf 1409} (2014) 074
  [arXiv:1406.6978 [hep-ph]].

\bibitem{Smirnov:2006bu}
  A.~Y.~Smirnov and R.~Zukanovich Funchal,
  Phys.\ Rev.\ D {\bf 74} (2006) 013001
  [hep-ph/0603009].

\bibitem{Hernandez:2014fha}
  P.~Hernandez, M.~Kekic and J.~Lopez-Pavon,
  Phys.\ Rev.\ D {\bf 90} (2014) no.6,  065033
  [arXiv:1406.2961 [hep-ph]].

\bibitem{Vincent:2014rja}
  A.~C.~Vincent, E.~F.~Martinez, P.~Hernandez, M.~Lattanzi and O.~Mena,
  JCAP {\bf 1504} (2015) no.04,  006
  [arXiv:1408.1956 [astro-ph.CO]].

 \bibitem{Abada:2015rta}
  A.~Abada, G.~Arcadi, V.~Domcke and M.~Lucente,
  ``Lepton number violation as a key to low-scale leptogenesis,''
  arXiv:1507.06215 [hep-ph].

\bibitem{Abada:2014zra}
  A.~Abada, G.~Arcadi and M.~Lucente,
  JCAP {\bf 1410} (2014) 001
  [arXiv:1406.6556 [hep-ph]].
  
 \bibitem{Abada:2014vea}
  A.~Abada and M.~Lucente,
  Nucl.\ Phys.\ B {\bf 885} (2014) 651
  [arXiv:1401.1507 [hep-ph]].

\bibitem{Asaka:2005pn}
  T.~Asaka and M.~Shaposhnikov,
  Phys.\ Lett.\ B {\bf 620} (2005) 17
  [hep-ph/0505013].

\bibitem{Shaposhnikov:2008pf}
  M.~Shaposhnikov,
  JHEP {\bf 0808} (2008) 008
  [arXiv:0804.4542 [hep-ph]].

\bibitem{Canetti:2012kh}
  L.~Canetti, M.~Drewes, T.~Frossard and M.~Shaposhnikov,
  Phys.\ Rev.\ D {\bf 87} (2013) 093006
  [arXiv:1208.4607 [hep-ph]].

   \bibitem{Casas:2001sr}
  J.~A.~Casas and A.~Ibarra,
  Nucl.\ Phys.\ B {\bf 618} (2001) 171
  [hep-ph/0103065].

\bibitem{Arganda:2015ija}
  E.~Arganda, M.~J.~Herrero, X.~Marcano and C.~Weiland,
  Phys.\ Lett.\ B {\bf 752} (2016) 46
  doi:10.1016/j.physletb.2015.11.013
  [arXiv:1508.05074 [hep-ph]].

\bibitem{Canetti:2013qna}
  L.~Canetti and M.~Shaposhnikov,
  Hyperfine Interact.\  {\bf 214} (2013) no.1-3,  5.

\bibitem{Gninenko:private}
S.~Gninenko, private communication.

\end{thebibliography}
}
\end{document}